\documentclass[pre,aps,reprint,amsmath]{revtex4-1}

\usepackage[T1]{fontenc} 
\usepackage{graphicx}
\usepackage[utf8]{inputenc}
\usepackage{lmodern}
\usepackage{bm}
\usepackage{bbold}
\usepackage[hidelinks]{hyperref}				
\usepackage[dvipsnames]{xcolor}
\usepackage{ulem}
\newcommand{\rv}{{\bf r}}
\newcommand{\Fv}{{\bf F}}
\newcommand{\fv}{{\bf f}}
\newcommand{\Jv}{{\bf J}}
\newcommand{\vel}{{\bf v}}
\newcommand{\nv}{{\bf n}}

\newcommand{\mv}{{\bf m}}
\newcommand{\ev}{{\bf e}}

\newcommand{\torque}{{\boldsymbol\tau}}
\newcommand{\Torque}{{\boldsymbol{\mathcal T}}}
\newcommand{\mysec}[2]{\section*{#1}\label{#2}}
\newcommand{\mysubsec}[2]{\subsection*{#1}\label{#2}}

\newcommand{\eps}{{\boldsymbol\epsilon}}
\newcommand{\ov}{{\boldsymbol\omega}}

\begin{document}

\title{Noether's Theorem in Statistical Mechanics}

\author{Sophie Hermann}
\email{Sophie.Hermann@uni-bayreuth.de}
\affiliation{Theoretische Physik II, Physikalisches Institut, 
  Universit{\"a}t Bayreuth, D-95447 Bayreuth, Germany}

\author{Matthias Schmidt}
\email{Matthias.Schmidt@uni-bayreuth.de}
\affiliation{Theoretische Physik II, Physikalisches Institut, 
  Universit{\"a}t Bayreuth, D-95447 Bayreuth, Germany}
\email{Matthias.Schmidt@uni-bayreuth.de}

\date{25 May 2021, Commun. Phys. \textbf{4}, 176 (2021).}

\begin{abstract}
Noether’s calculus of invariant variations yields exact identities from functional symmetries. The standard application to an action integral allows to identify conservation laws. Here we rather consider generating functionals, such as the free energy and the power functional, for equilibrium and driven many-body systems. Translational and rotational symmetry operations yield mechanical laws. These global identities express vanishing of total internal and total external forces and torques. We show that functional differentiation then leads to hierarchies of local sum rules that interrelate density correlators as well as static and time direct correlation functions, including memory. For anisotropic particles, orbital and spin motion become systematically coupled. The theory allows us to shed new light on the spatio-temporal coupling of correlations in complex systems. As applications we consider active Brownian particles, where the theory clarifies the role of interfacial forces in motility-induced phase separation. For active sedimentation, the center-of-mass motion is constrained by an internal Noether sum rule.
\end{abstract}

\maketitle

\mysec{Introduction}{SECintroduction}
Emmy Noether's 1918 Theorems for 
{\it Invariant Variation Problems}
\cite{noether1918,neuenschwander2011}, as applied to action
functionals both in particle-based and field-theoretic contexts, form
a staple of our fundamental description of nature.  The formulation of
energy conservation in general relativity had been the then open and
vexing problem, that triggered Hilbert and Klein to draw Noether into
their circle, and she ultimately solved the problem \cite{byers1998}.
Her deep insights into the relationship of the emergence and validity
of conservation laws with the underlying local and global symmetries
of the system has been exploited for over a century. 

While Noether's work has been motivated by the then ongoing
developments in general relativity, being a mathematician, she has
formulated her theory in a much broader setting than given by the
specific structure of the action as a space-time integral over a
Lagrangian density, as formulated by Hilbert in 1916 for Einstein's
field equations. Her work rather applies to functionals of a much more
general nature, with only mild assumptions of analyticity and careful
treatment of boundary conditions of integration domains.

In Statistical Physics, the use of Noether's Theorems is significantly
more scarce, as opposed to both classical mechanics and high energy
physics. Notable exceptions include the square-gradient treatment of
the free gas-liquid interface, cf.\ Rowlinson and Widom's enlightening
description \cite{rowlinson2002} of van der Waals' prototypical
solution \cite{vanderwaals1893}. In a striking analogy, the square gradient contribution to
the free energy is mapped onto kinetic energy of an effective particle
that traverses in time between two potential energy maxima of equal
height. Exploiting energy conservation in the effective system yields
a first integral, which constitutes a nontrivial identity in the
statistical problem. This reasoning has been generalized to the
delicate problem of the three-phase contact line that occurs at a
triple point of a fluid mixture \cite{kerins1983,bukman2003}. While
these treatments strongly rely on the square-gradient approximation,
Boiteux and Kerins also developed a method that they refer to as
variation under extension, which permitted them to treat more general
cases \cite{boiteux1983}.

Evans has derived a number of exact sum rules for inhomogeneous fluids
in his pivotal treatment of the field \cite{evans1979}. While not
spelling out any connection to Noether's work, he carefully examines
the effects of spatial displacements on distribution functions. This
shifting enables him, as well as Lovett et
al.\ \cite{lovett1976} and Wertheim \cite{wertheim1976} in earlier
work, to identify systematically the effects that result from the
displacement and formulate these as highly nontrivial interrelations
(``sum rules'') between correlation functions. This approach was
subsequently generalized to higher than two-body direct
\cite{kayser1980} and density \cite{baus1984} correlation functions
and the relationship to integral equation theory was addressed
\cite{lovett1991,baus1992}.
Considering also rotations Tarazona and Evans \cite{tarazona1983} have
addressed the case of anisotropic particles, where their sum rules
correct earlier results by Gubbins \cite{gubbins1980}.
The exploitation of the fundamental spatial symmetries
\cite{lovett1976,wertheim1976,evans1979,kayser1980,tarazona1983,
  baus1984,lovett1991,baus1992} appears to be intimately
related to Noether's thinking. This is no coincidence, as Evans'
classical density functional approach (DFT) is variational as is the
general problem that she addresses.

DFT constitutes a powerful modern framework for the description of a
broad range of interfacial, adsorption, solvation, and phase
phenomenology in complex systems
\cite{evans1979,hansen2013,evans2016specialIssue,roth2010review}. Examples
of recent pivotal applications include the treatments of
hydrophobicity \cite{levesque2012jcp,jeanmairet2013jcp,evans2015prl,
  evans2019pnas,rensing2019pnas,evans2016prl,chacko2017} and of drying
\cite{evans2019pnas,evans2015prl,evans2016prl}, electrolytes near
surfaces \cite{martinjimenez2017natCom}, dense fluid structuring as
revealed in atomic force microscopy \cite{hernandez-munoz2019},
thermal resistance of liquid-vapor interfaces \cite{muscatello2017},
and layered freezing in confined colloids \cite{xu2008}.  Xu and Rice
\cite{xu2008} have used the sum rules of Lovett, Mou, and Buff \cite{lovett1976} and  Wertheim \cite{wertheim1976} (LMBW) to carry out a bifurcation
analysis of the confined fluid state.  The sum rules were instrumental
for investigating a range of topics, such as precursors to freezing
\cite{brader2008precursor}, nonideal \cite{walz2010} and cluster
crystals \cite{haering2015}, liquid crystal deformations
\cite{haering2020phd}, and --prominently-- interfaces of liquids
\cite{bryk1997,henderson1984,tejero1985,iatsevitch1997,kasch1993}. A
range of further techniques besides DFT was used in this context,
including integral equation theory
\cite{bryk1997,brader2008precursor,iatsevitch1997}, mode-coupling
theory~\cite{mandal2017}, and Mori-Zwanzig equations
\cite{walz2010,haering2020phd}.

Much of very current attention in Statistical Physics is devoted to
nonequilibrium and active systems that are driven in a controlled way
out of equilibrium, such as e.g.\ active Brownian particles
\cite{farage2015,paliwal2018,paliwal2017activeLJ} and 
magnetically controlled topological transport of colloids
\cite{loehr2018natComm,loehr2018commsPhys,rossi2019}.
The power functional (variational) theory \cite{schmidt2013pft} (PFT)
offers to obtain a unifying perspective on nonequilibrium problems
such as the above.  In PFT the (time-dependent) density distribution
is complemented by the (time-dependent) current distribution as a
further variational field. A rigorous extremal principle determines
the motion of the system, on the one-body level of correlation
functions. The concept enabled to obtain a fundamental understanding
and quantitative description of a significant array of nonequilibrium
phenomena, such as the identification of superadiabatic forces
\cite{fortini2014prl}, the treatment of active Brownian particles
\cite{krinninger2016prl,hermann2019prl,hermann2019pre,hermann2020polarization},
of viscous \cite{delasheras2018velocityGradient}, structural
\cite{stuhlmueller2018prl,delasheras2020prl} and flow forces
\cite{delasheras2020prl}.
Crucially, the DFT remains relevant for the description of
nonequilibrium situations, via the adiabatic construction
\cite{schmidt2013pft,fortini2014prl}, which captures those parts of
the dynamics that functionally depend on the density distribution
alone, and do so instantaneously. 
Both equilibrium DFT and nonequilibrium PFT provide formally exact
variational descriptions of their respective realm of Statistical
Physics. While action integrals feature in neither formulation, the
relevant functionals do fall into the general class of functionals
that Noether considered in her work. 

\begin{figure}
  \includegraphics[width=0.79\columnwidth,angle=0]{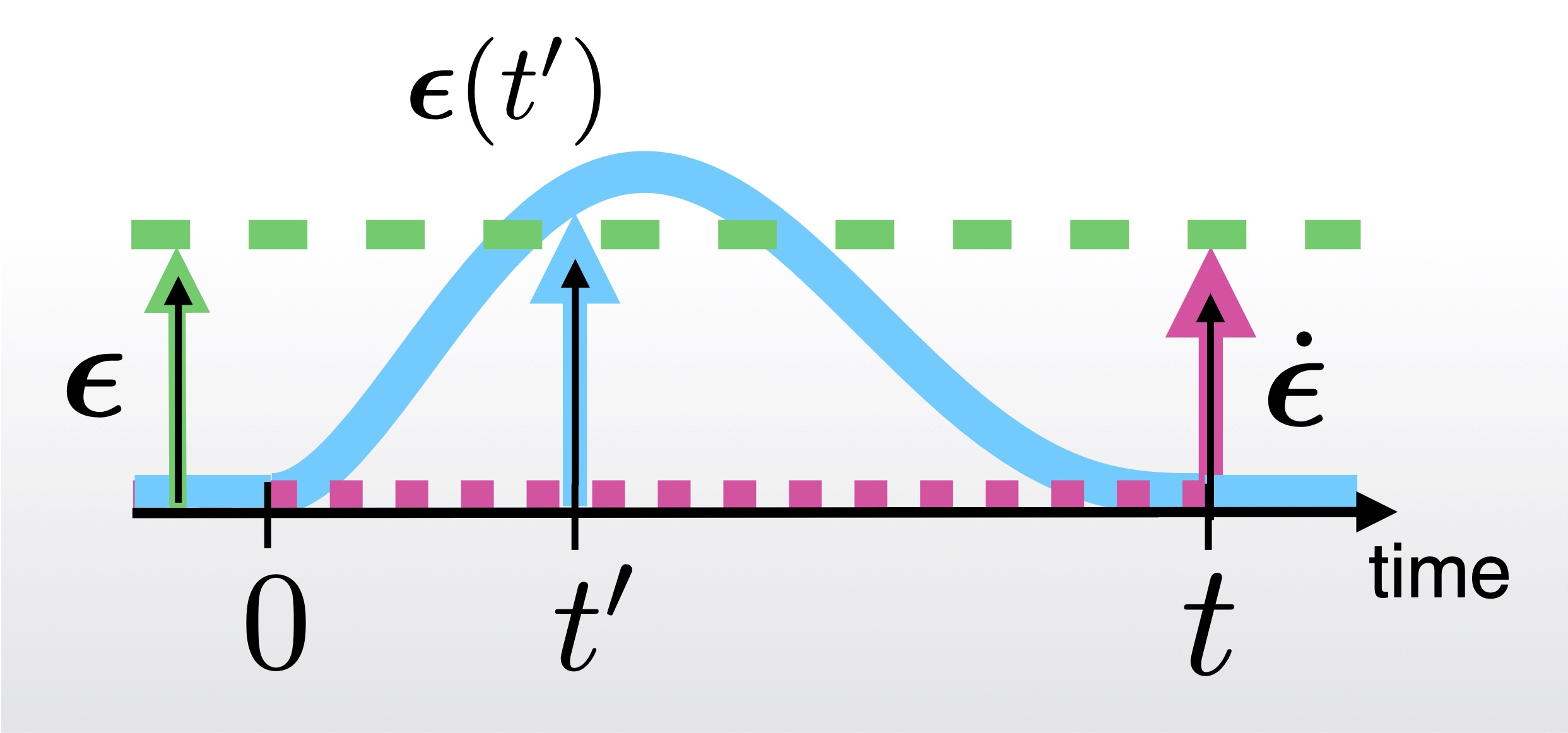}
  \caption{\textbf{Illustration of the three types of dynamical
    transformations considered.} The system is spatially displaced by $\eps=\rm
    const$ at all times (green dashed), analogously to the operation
    in equilibrium. The system is dynamically displaced by $\eps(t')$,
    such that the spatial displacement vanishes at the boundaries of the
    considered time interval, $\eps(0)=\eps(t)=0$ (cyan solid). The
    system is displaced instantaneously only at the latest time $t$, such that the differential displacement is
    $\dot\eps dt$ (purple dotted).}
  \label{fig1}
\end{figure}

Here we apply Noether's Theorem
to Statistical Physics.  We first
introduce the basic concepts via treating spatial translations for both the
partition sum and for the free energy density functional. 
 Crucially, considering the symmetries of the partition sum does not require to engage with density functional concepts; the elementary definition suffices. We demonstrate that this approach is
consistent with the earlier work in equilibrium
\cite{lovett1976,wertheim1976, evans1979,kayser1980,tarazona1983,
  baus1984,lovett1991,baus1992}, and that it enables one to go, with
relative ease, beyond the sum rules that these authors formulated. In
nonequilibrium, we apply the same symmetry operations to the
time-dependent case and obtain novel exact and nontrivial identities
that apply for driven and active fluids.  The three different types of
time-dependent shifting are illustrated in Fig.~\ref{fig1}.  The
resulting sum rules are different from the nonequilibrium
Ornstein-Zernike (NOZ) relations \cite{brader2013noz,brader2014noz},
but they possess an equally fundamental status.  
 We also consider the more general case of anisotropic interparticle interactions and treat rotational invariance both in and out of equilibrium.
 To illustrate the theory we apply it to both passive and active phase coexistence as well as to active sedimentation under gravity.

\begin{figure}
  \includegraphics[width=0.59\columnwidth,angle=0]{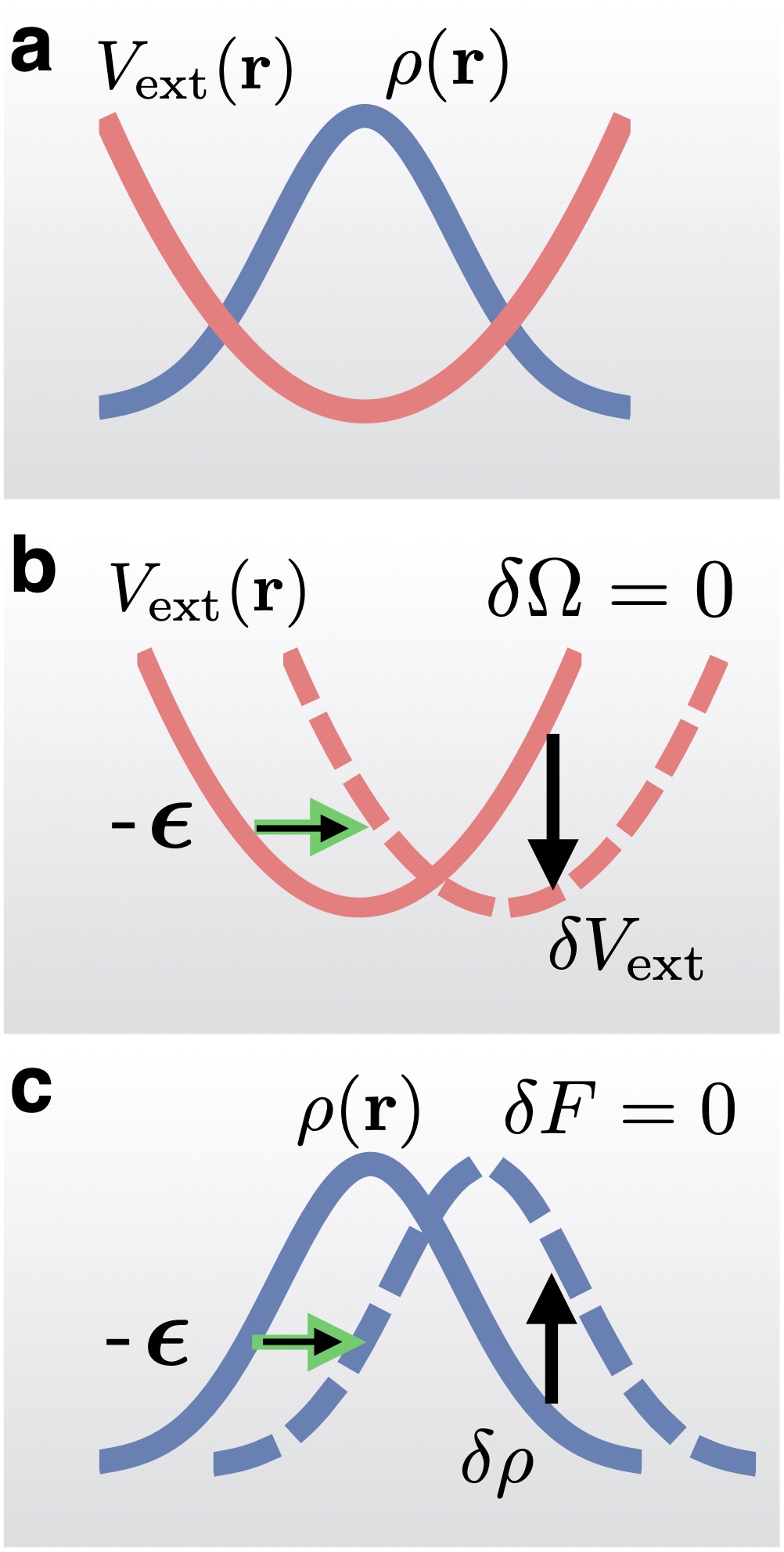}
  \caption{\textbf{Illustrations of the effects induced by shifting in
    equilibrium.} \textbf{a}~In the presence of external potential $V_{\rm
      ext}(\rv)$, the system develops an inhomogeneous density profile
    $\rho(\rv)$, where $\textbf{r}$ denotes the position coordinate. \textbf{b}~Shifting the external potential by a displacement
    vector $-\eps$ (green arrow) induces a local change in external potential $\delta
    V_{\rm ext}(\rv)$ (black arrow) between the original (solid line) and the shifted external potential (dashed line); the grand potential is invariant,
    $\delta\Omega=0$. \textbf{c}~The displaced density profile  (dashed line) implies a
    local change $\delta\rho(\rv)$ (black arrow) in comparison to the initial density profile (solid line), which leaves the intrinsic free
    energy unchanged, $\delta F=0$.}
  \label{fig2}
\end{figure}

\mysec{Results and Discussion}{SECResults}
\mysubsec{Adiabatic state}{SECadiabaticState}We start with an initial
illustration of Noether's concept as applied to the grand potential $\Omega$. We consider spatial
translations of the position coordinate $\textbf{r}$ at fixed chemical 
potential $\mu$ and fixed temperature $T$. The system is under the
influence of a one-body external potential $V_{\rm ext}(\rv)$, 
cf.\ Fig.~\ref{fig2}(a). We take $V_{\rm ext}(\rv)$ to also
describe container walls, such that there is no need for the system
volume as a further thermodynamic variable. For the moment we only examine systems completely bounded by external walls. Systems with open boundaries are considered below. 
Clearly the value of the grand potential 
$\Omega$ is independent of the global location of a system. 
Hence spatial shifting by a (global) displacement vector $\eps$ leaves the
value of $\Omega$ invariant. 
To exploit this symmetry in a variational
setting, note that the value of $\Omega$ depends on the function
$V_{\rm ext}(\rv)$, hence $V_{\rm ext}(\rv)\to\Omega$ constitutes a
functional map, at given $\mu$ and~$T$. Here the grand potential is
defined by its elementary Statistical Mechanics form $\Omega[V_{\rm
    ext}]=-k_\text{B}T\ln\Xi$, with the grand partition sum $\Xi$ depending
functionally via the Boltzmann factor on $V_{\rm ext}(\rv)$. The
spatial displacement amounts to the operation $V_{\rm ext}(\rv)\to
V_{\rm ext}(\rv+\eps)$, cf.\ Fig.~\ref{fig2}(b).  For small
$\eps$ we can Taylor expand to linear order: $V_{\rm
  ext}(\rv+\eps)=V_{\rm ext}(\rv)+\delta V_{\rm ext}(\rv)$, where
$\delta V_{\rm ext}(\rv)=\eps\cdot\nabla V_{\rm ext}(\rv)$ indicates
the local change of the external potential that is induced by the
shift. As $\Omega[V_{\rm ext}]$ is invariant under the shift (which
can be shown by translating all particle coordinates in $\Xi$
accordingly), we have
\begin{align}
  \Omega[V_{\rm ext}] &= \Omega[V_{\rm ext}+\delta V_{\rm ext}]
  \notag\\
  &=
  \Omega[V_{\rm ext}] + \int d\rv 
  \frac{\delta \Omega[V_{\rm ext}]}{\delta V_{\rm ext}(\rv)}
  \eps \cdot \nabla V_{\rm ext}(\rv).
  \label{EQOmegaShifted}
\end{align}
Here the second equality constitutes a functional Taylor expansion in
$\delta V_{\rm ext}(\rv)$ to linear order, and $\delta \Omega[V_{\rm
    ext}]/\delta V_{\rm ext}(\rv)$ indicates the functional derivative
of $\Omega[V_{\rm ext}]$ with respect to its argument, evaluated here
at the unshifted function $V_{\rm ext}(\rv)$, i.e.\ $\eps=0$.
It is a straightforward elementary exercise
\cite{evans1979,hansen2013} to show via explicit calculation that
$\delta\Omega[V_{\rm ext}]/\delta V_{\rm ext}(\rv)=\rho(\rv)$, where
$\rho(\rv)=\langle\sum_i\delta(\rv-\rv_i)\rangle_{\rm eq}$ is the
microscopically resolved one-body density profile. Here $\rv_i$
indicates the position of particle $i=1\ldots N$, with $N$ being the
total number of particles, $\delta(\cdot)$ indicates the Dirac
distribution, and the average is over the equilibrium distribution at
fixed $\mu$ and~$T$; the sum runs over all particles $i=1\ldots N$.

Comparing the left and right hand sides of \eqref{EQOmegaShifted} and
noticing that $\eps$ is arbitrary, we conclude
\begin{align}
  \Fv_{\rm ext}^{\rm tot} &=
  -\int d\rv \rho(\rv) \nabla V_{\rm ext}(\rv) =0,
  \label{EQforceExternalTotal}
\end{align}
where we have defined the total external force $\Fv_{\rm ext}^{\rm
  tot}$ using the one-body fields $\rho(\rv)$ and $V_{\rm
  ext}(\rv)$. It is straightforward to show the equivalence with the
more elementary form $\Fv_{\rm ext}^{\rm tot} = -\langle \sum_i
\nabla_i V_{\rm ext}(\rv_i) \rangle_{\rm eq}$, where $\nabla_i$
indicates the derivative with respect to $\rv_i$. Clearly,
\eqref{EQforceExternalTotal} expresses the vanishing of the total
external force. (Consider e.g.\ the gravitational weight of an
equilibrium colloidal sediment being balanced by the force that the
lower container wall exerts on the particles.)

Equation \eqref{EQforceExternalTotal} was previously obtained by Baus
\cite{baus1984}. Here we have identified it as a Noether sum rule for
the case of spatial displacement of $\Omega[V_{\rm ext}]$.
We can generate local sum rules by observing that
\eqref{EQforceExternalTotal} holds for any form of $V_{\rm ext}(\rv)$
and that hence $V_{\rm ext}(\rv)\to\rho(\rv)$ constitutes a functional
map (defined by the grand canonical average
$\langle\hat\rho(\rv)\rangle_{\rm eq}$, which features $V_{\rm
  ext}(\rv)$ in the equilibrium many-body probability
distribution). We hence functionally differentiate
\eqref{EQforceExternalTotal} by $V_{\rm ext}(\rv')$, where $\rv'$ is a
new position variable. The first and the $n$th functional derivatives
yield, respectively, the identities
\begin{align}
  \nabla\rho(\rv) &= 
  -\int d\rv' \beta H_2(\rv,\rv') \nabla' V_{\rm ext}(\rv'),
  \label{EQBobH2}\\
  \sum_{\alpha=1}^n \nabla_\alpha H_n &= -
  \int d\rv_{n+1} \beta V_{\rm ext}(\rv_{n+1}) \nabla_{n+1} H_{n+1},
  \label{EQBobHnHierarchy}
\end{align}
where $\beta=1/(k_\text{B}T)$, with $k_\text{B}$ indicating the Boltzmann constant,
$H_2(\rv,\rv')=-\delta\rho(\rv)/\delta \beta V_{\rm ext}(\rv')$ is the
two-body correlation function of density fluctuations, and $H_n=\delta
H_{n-1}/\delta \beta V_{\rm ext}(\rv_n)$ is its $n$-body version
\cite{evans1979,hansen2013}. Here position arguments have been omitted
for clarity: $H_n\equiv H_n(\rv_1\ldots\rv_n)$, and $\nabla_\alpha$
indicates the derivative with respect to $\rv_\alpha$. The variable
names $\rv$ and $\rv'$ have been interchanged in \eqref{EQBobH2} and
$\nabla'$ indicates the derivative with respect to $\rv'$. The
derivation of \eqref{EQBobH2} and \eqref{EQBobHnHierarchy} requires
spatial integration by parts. Recall that boundary terms vanish
 as we only consider systems with impenetrable bounding walls. 

The sum rule \eqref{EQBobH2} has been obtained by LMBW \cite{lovett1976,wertheim1976} and by Evans \cite{evans1979} on
the basis of shifting considerations. The present formulation based on
Noether's more general perspective allows to reproduce \eqref{EQBobH2}
with great ease and to generalize to the hierarchy
\eqref{EQBobHnHierarchy}, as previously obtained by Baus
\cite{baus1984}. Equation \eqref{EQBobH2} has the interpretation of
the density gradient $\nabla\rho(\rv)$ being stabilized by the action
of the external force field, $-\nabla V_{\rm ext}(\rv)$. The effect is
mediated by $\beta H_2(\rv,\rv')$, where the correlation of the
density fluctuations is due to the coupled nature of the interparticle
interactions. Equation \eqref{EQBobHnHierarchy} is the multi-body
generalization of this mechanism. Via multiplying \eqref{EQBobH2} by
$V_{\rm ext}(\rv)$, integrating over $\rv$, and using
\eqref{EQforceExternalTotal}, and iteratively repeating this process
for all orders, one obtains a multi-body analogue of the vanishing
external force \eqref{EQforceExternalTotal}: 
\begin{align}
\int d\rv_1 V_{\rm
  ext}(\rv_1)\ldots \int d\rv_n V_{\rm ext}(\rv_n)\nabla_\alpha
H_n=0,
\label{EQHn}
\end{align} 
for $\alpha=1\ldots n$.

We turn to intrinsic contributions. As Noether's Theorem poses no
restriction on the type of physical functional, we consider the
intrinsic Helmholtz free energy $F[\rho]$ as a functional of the
density profile as its natural argument. Here a functional Legendre
transform \cite{hansen2013,evans1979} yields $F[\rho]=\Omega[V_{\rm
    ext}]-\int d\rv (V_{\rm ext}(\rv)-\mu)\rho(\rv)$. Crucially,
$F[\rho]$ is independent of $V_{\rm ext}(\rv)$, and its excess (over
ideal gas) contribution $F_{\rm exc}[\rho]$ is specific to the form of
the interparticle interaction potential $u(\rv^N)$; here we use the
shorthand $\rv_1\ldots\rv_N\equiv \rv^N$.  The full intrinsic free
energy functional consists of a sum of ideal gas and excess
contributions, i.e.\ $F[\rho]=k_\text{B}T\int d\rv
\rho(\rv)[\ln(\rho(\rv)\Lambda^D)-1]+F_{\rm exc}[\rho]$, where
$\Lambda$ is the (irrelevant) thermal de Broglie wavelength and $D$ is
the dimensionality of space.

As $u(\rv^N)$ is globally translationally invariant, $F_{\rm exc}[\rho]$ will
not change its value when evaluated at a spatially displaced density,
$\rho(\rv+\eps)=\rho(\rv)+\delta \rho(\rv)$, where
$\delta\rho(\rv)=\eps\cdot\nabla\rho(\rv)$,
cf.\ Fig.~\ref{fig2}(c). Hence in analogy to
\eqref{EQOmegaShifted}, we obtain $F_{\rm exc}[\rho] = F_{\rm
  exc}[\rho+\delta\rho] = F_{\rm exc}[\rho] + \int d\rv (\delta F_{\rm
  exc}[\rho]/\delta\rho(\rv)) \eps\cdot\nabla\rho(\rv)$.  Again $\eps$
is arbitrary. As boundary terms vanish in the considered systems, integration by parts yields
\begin{align}
  \Fv_{\rm ad}^{\rm tot} &=
  -\int d\rv \rho(\rv) \nabla 
  \frac{\delta F_{\rm exc}[\rho]}{\delta\rho(\rv)}
  = \int d\rv \rho(\rv)\fv_{\rm ad}(\rv) = 0,
  \label{EQforceInternalTotal}
\end{align}
where the first equality expresses the total internal force $\Fv_{\rm
  ad}^{\rm tot}=-\langle \sum_i \nabla_i u(\rv^N)\rangle_{\rm eq}$ in
DFT language. Hence \eqref{EQforceInternalTotal} expresses the fact
that the total internal force vanishes in equilibrium; the more
general time-dependent case is treated below.
The functional derivatives of $F_{\rm exc}[\rho]$
constitute direct correlation functions
\cite{ornstein1914,hansen2013,evans1979}, with the lowest order being
the one-body direct correlation function $c_1(\rv) = -\delta \beta
F_{\rm exc}[\rho]/\delta\rho(\rv)$. The equilibrium (``adiabatic'')
force field is simply $\fv_{\rm ad}(\rv) = k_\text{B}T\nabla c_1(\rv)$. This
one-body force field arises from the interparticle forces that all
other particles exert on the particle that resides at position $\rv$.

From the global internal Noether sum rule \eqref{EQforceInternalTotal}, we can
obtain local sum rules by observing that \eqref{EQforceInternalTotal}
holds for all $\rho(\rv)$ and hence that its functional derivative
with respect to $\rho(\rv)$ vanishes identically, i.e.,
\begin{align}
  \nabla c_1(\rv) &= \int d\rv' c_2(\rv,\rv') \nabla' \rho(\rv'),
  \label{EQBobDirectTwoBody}\\
  \sum_{\alpha=1}^n \nabla_\alpha c_n
  &=-\int d\rv_{n+1}\rho(\rv_{n+1})
  \nabla_{n+1}c_{n+1},
  \label{EQBobDirectHierarchy}
\end{align}
where $c_2(\rv,\rv')$ is the (inhomogeneous) two-body direct
correlation function of liquid state theory \cite{hansen2013};
$c_n\equiv c_n(\rv_1\ldots\rv_n)$ is the $n$-body direct correlation
function, defined recursively via $c_{n+1}=\delta
c_n/\delta\rho(\rv_{n+1})$. As identified by LMBW \cite{lovett1976,wertheim1976} and Evans \cite{evans1979},
\eqref{EQBobDirectTwoBody} expresses the conversion of the density
gradient, via the two-body direct correlations, to the locally
resolved intrinsic force field; recall that $\fv_{\rm
  ad}(\rv)=k_\text{B}T\nabla c_1(\rv)$. Via the Noether formalism the
corresponding hierarchy \eqref{EQBobDirectHierarchy} is obtained
straightforwardly from repeated functional differentiation
\cite{kayser1980,baus1984,lovett1991} with respect to $\rho(\rv)$.
Note that similar to the structure of
\eqref{EQBobHnHierarchy}, only consecutive terms of order $n$ and
$n+1$ are directly coupled in \eqref{EQBobDirectHierarchy}.  A
multi-body version of \eqref{EQforceInternalTotal} is obtained by
multiplying \eqref{EQBobDirectTwoBody} with $\rho(\rv)$, integrating
over~$\rv$, exploiting \eqref{EQforceInternalTotal}, and iterating for
all orders. The result is: 
\begin{align}
\int d\rv_1\rho(\rv_1)\ldots\int
d\rv_n\rho(\rv_n)\nabla_\alpha c_n=0, \label{EQcn}
\end{align}
for $\alpha=1\ldots n$. In the
case $\alpha=n=1$ we recover \eqref{EQforceInternalTotal}.

The global sum rule \eqref{EQforceInternalTotal} of vanishing total
internal force can be straightforwardly obtained by more elementary
analysis. We exploit translation invariance in this non-functional
setting: $u(\rv^N)\equiv u(\rv_1+\eps\ldots\rv_N+\eps)$.  Then the
derivative with respect to $\eps$ vanishes, $0=\partial
u(\rv_1+\eps\ldots\rv_N+\eps)/\partial \eps =\sum_i \nabla_i
u(\rv^N)$. The latter expression follows from the chain rule and
constitutes the total internal force (up to a minus sign), which hence
vanishes for each microstate $\rv^N$. The connection to (the
many-body version of) Newton's third law {\it actio equals reactio}
becomes apparent in the rewritten form $-\nabla_\alpha
u(\rv^N)=\sum_{i\neq\alpha}\nabla_i u(\rv^N)$, for $\alpha=1\ldots
N$. The thermal equilibrium average is then trivial and on average
$\Fv_{\rm ad}^{\rm tot}=0$.  This argument is very general and it
remains true if the average is taken over a nonequilibrium
many-body distribution function.  The total internal force in such a
general situation is
\begin{align}
  \Fv_{\rm int}^{\rm tot}=-\Big\langle\sum_i\nabla_i
  u(\rv^N)\Big\rangle=0, \label{EQFint}
\end{align}
where the average is taken over the nonequilibrium many-body
probability distribution at time $t$. We have hence proven that the
total internal force vanishes for all times $t$.  In addition, the
particles can possess additional degrees of freedom $\ov_i$,
$i=1\ldots N$, as is the case for the orientation vectors of active
Brownian particles, to which we return after first laying out the setup in nonequilibrium.

\mysubsec{Nonequilibrium states}{SECnonequilibrium}To be specific, we
consider overdamped Brownian motion, at constant temperature $T$ and
with no hydrodynamic interactions present \cite{hansen2013}, as
described by the Smoluchowski (Fokker-Planck) equation. The
microscopically resolved local internal force field is $\fv_{\rm
  int}(\rv,t)=-\langle\sum_i\delta(\rv-\rv_i)\nabla_i
u(\rv^N)\rangle/\rho(\rv,t)$, where the average is over the
nonequilibrium distribution (which evolves in time according to the
Smoluchowski equation) at time~$t$. The total internal force is then
the spatial integral $\Fv_{\rm int}^{\rm tot}=\int
d\rv\rho(\rv,t)\fv_{\rm int}(\rv,t)$. Applying Noether's Theorem to
the nonequilibrium case requires to have a variational description, as
is provided by PFT \cite{schmidt2013pft}. Here the variational fields
are the time-dependent density profile $\rho(\rv,t)$ and the
time-dependence one-body current
$\Jv(\rv,t)=\langle\sum_i\delta(\rv-\rv_i)\vel_i\rangle$, where
$\vel_i(\rv^N,t)$ is the configurational velocity of particle $i$. The
microscopically resolved average velocity profile is
$\vel(\rv,t)=\Jv(\rv,t)/\rho(\rv,t)$. PFT ascertains the splitting
$\fv_{\rm int}(\rv,t)=\fv_{\rm ad}(\rv,t)+\fv_{\rm sup}(\rv,t)$, where
the adiabatic force field is that in a corresponding equilibrium
(``adiabatic'') system with identical instantaneous density profile,
$\fv_{\rm ad}(\rv,t)=-\nabla\delta F_{\rm
  exc}[\rho]/\delta\rho(\rv,t)$ and $\fv_{\rm sup}(\rv,t)$ is the
superadiabatic internal force field, obtained as $\fv_{\rm
  sup}(\rv,t)=- \delta P_t^{\rm exc}[\rho,\Jv]/ \delta\Jv(\rv,t)$,
where $P_t^{\rm exc}[\rho,\Jv]$ is the superadiabatic excess free
power functional \cite{schmidt2013pft}.

\begin{figure}
  \includegraphics[width=0.59\columnwidth,angle=0]{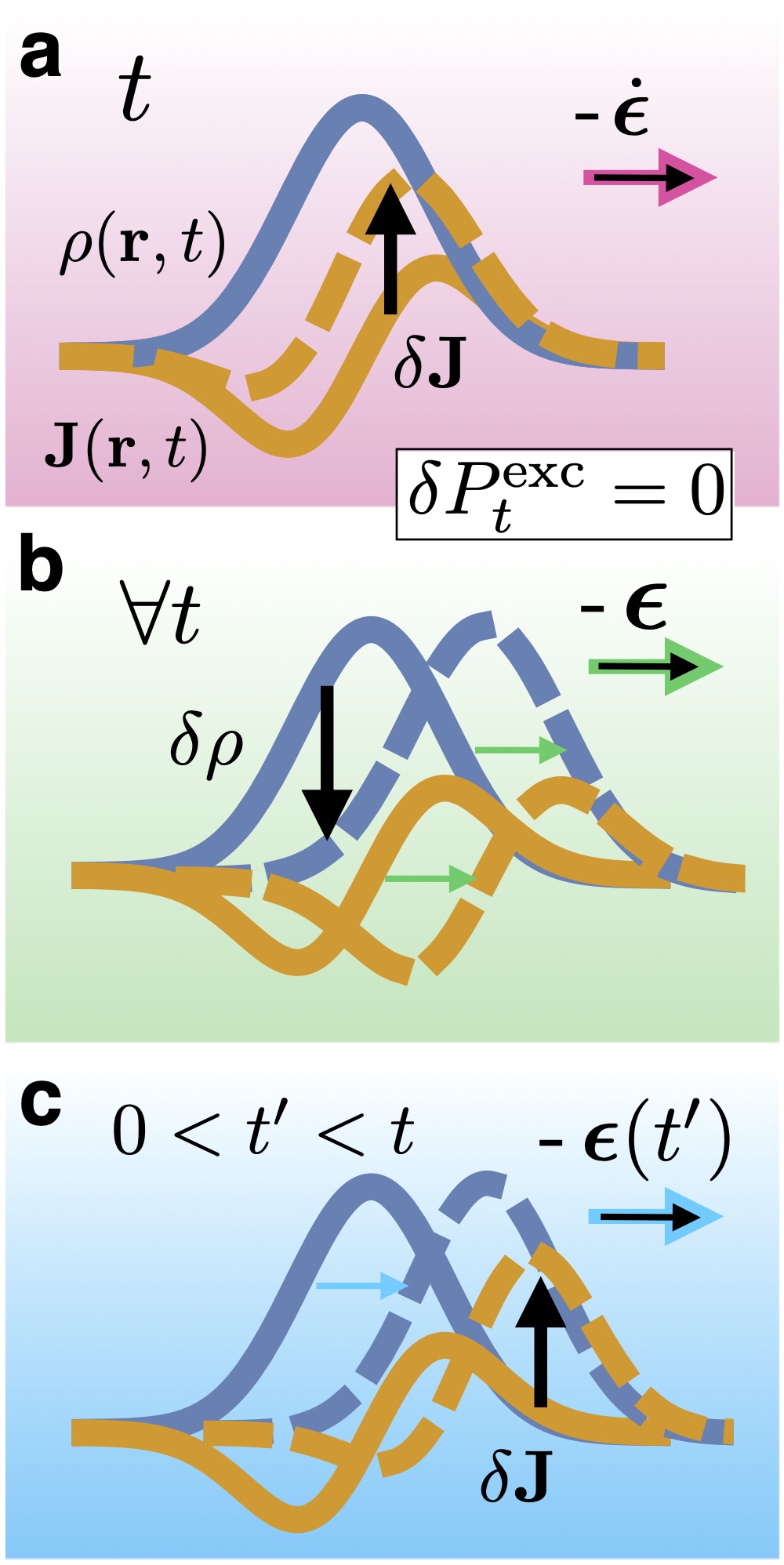}
  \caption{\textbf{Illustrations of the effects induced by shifting in nonequilibrium.   
  } All transformations affect both the density profile $\rho(\rv,t)$ (blue) and the current profile $\Jv(\rv,t)$ (yellow), while
    the superadiabatic excess power functional is invariant, $\delta
    P_t^{\rm exc}=0$. Here $t$ indicates the time and $\textbf{r}$ is the position coordinate.  
    \textbf{a}~An instantaneous spatial shift by
    $\dot\eps$ at time $t$ induces a current change $\delta\Jv=-\dot
    \eps\rho$. \textbf{b}~A static shift by $\eps=\rm const$ is applied at
    all times $t$. \textbf{c}~A time-dependent shift $\eps(t')$ is applied
    between inital time 0 and final time $t$ of the considered time
    interval.  }
  \label{fig3}
\end{figure}

Crucially, $\fv_{\rm ad}(\rv,t)$ is a density functional, independent
of the flow in the system, while $\fv_{\rm sup}(\rv,t)$ is a kinematic
functional, i.e.\ with dependence on both $\rho(\rv,t)$ and
$\Jv(\rv,t)$, including memory, i.e.\ dependence on the value of the
fields at times $<t$. As the local force fields split into adiabatic
and superadiabatic contributions, so do the total forces: $\Fv_{\rm
  int}^{\rm tot}=\int d\rv\rho\fv_{\rm int}=\int d\rv\rho\fv_{\rm
  ad}+\int d\rv\rho\fv_{\rm sup}\equiv \Fv_{\rm ad}^{\rm tot}+\Fv_{\rm
  sup}^{\rm tot}$. We have seen above that $\Fv_{\rm int}^{\rm
  tot}=\Fv_{\rm ad}^{\rm tot}=0$. Hence also
\begin{align}
  \Fv_{\rm sup}^{\rm tot} &= -\int d\rv\rho(\rv,t)
  \frac{\delta P_t^{\rm exc}[\rho,\Jv]}{\delta\Jv(\rv,t)}
  \equiv\int d\rv \rho\fv_{\rm sup}=0.
  \label{EQforceSuperadiabaticTotal}
\end{align}
While the above reasoning required to rely on the many-body level, the
same result \eqref{EQforceSuperadiabaticTotal} can be
straightforwardly obtained in a pure Noetherian way, by considering an
instantaneous shift of coordinates at time $t$,
i.e.\ $\Jv(\rv,t)\to\Jv(\rv,t)-\dot\eps\rho(\rv,t)$,
cf.\ Fig.~\ref{fig3}(a).
(Fig.~\ref{fig3} gives an overview of the three different types of shifting.)
 Here $\dot\eps$ is the
corresponding instantaneous change in velocity with
$\vel(\rv,t)\to\vel(\rv,t)-\dot\eps$, as obtained by dividing the
current by the density profile.  Due to the overdamped nature of the
Smoluchowski dynamics, the internal interactions are unaffected and
the shift constitutes a symmetry operation for the generator of the
superadiabatic forces, $P_t^{\rm exc}[\rho,\Jv]$. Hence the
instantaneous current perturbation
$\delta\Jv(\rv,t)=-\dot\eps\rho(\rv,t)$ that is generated by the
invariance transformation leads to $P_t^{\rm exc}[\rho,\Jv]=P_t^{\rm
  exc}[\rho,\Jv+\delta\Jv]= P_t^{\rm exc}[\rho,\Jv] - \int d\rv
(\delta P_t^{\rm exc}[\rho,\Jv]/\delta
\Jv(\rv,t))\cdot\dot\eps\rho(\rv,t)$. As $\dot\eps$ is arbitrary, we
obtain \eqref{EQforceSuperadiabaticTotal}. Treating the dynamical
adiabatic contribution $\dot F[\rho]=\int d\rv\Jv(\rv,t)\cdot\nabla
\delta F_{\rm exc}[\rho]/\delta\rho(\rv,t)$ in the same way, we
re-obtain \eqref{EQforceInternalTotal}.
As Noether's theorem is converse, it allows for alternative reasoning:
the invariance of $P_t^{\rm exc}[\rho,\Jv]$ to the instantaneous shift
of the current (or analogously of the velocity) can hence be 
  derived from~\eqref{EQforceSuperadiabaticTotal} by simply
reversing the above chain of arguments.

We can generate nonequilibrium sum rules by differentiating the
Noether identity \eqref{EQforceSuperadiabaticTotal} with respect to
$\Jv(\rv,t)$, which yields
\begin{align}
  \int d\rv' {\sf M}_2  (\rv,\rv',t)\rho(\rv',t) &= 0,
  \label{EQM2identity}
\end{align}
where ${\sf M}_2(\rv,\rv',t)=-\beta\delta^2 P_t^{\rm
  exc}[\rho,\Jv]/\delta\Jv(\rv,t)\delta\Jv(\rv',t)$ is the tensorial
two-body equal-time direct correlation function
\cite{brader2013noz,brader2014noz}. Its $n$-body version is obtained
from ${\sf M}_{n+1}(\rv_1\ldots\rv_{n+1},t)=\delta {\sf
  M}_n(\rv_1\ldots\rv_n,t)/\delta\Jv(\rv_{n+1},t)$, and it satisfies
the hierarchy 
\begin{align}
\int d\rv_{n+1} {\sf M}_{n+1} \rho(\rv_{n+1},t) = 0, \label{EQMnidentity}
\end{align} 
as obtained by differentiating \eqref{EQM2identity} repeatedly with
respect to the current.

\mysubsec{Open boundaries}{SECopenBoundaries} 
All our considerations have
been based on applying the symmetry operation to the entire system  confined by external walls.
The effects of these system walls are modelled by a suitable form of
$V_\text{ext}(\rv)$.  The position integrals formally run over all
space, with the cutoff provided by hard (or steeply rising) external
wall potentials. In many practical and relevant situations, it is
more useful to consider a system with open boundaries.  Alternatively one can consider only a subvolume~$V$ of the entire system, and restrict the accounting of force contributions to those particles
that reside inside of $V$ at a given time. 
In doing so, one needs to
take account of boundary effects \cite{requardt1988}, as the
boundaries of $V$ are open, such that interparticle forces can be
transmitted, and flow can occur. 

In case that there are no net boundary contributions, all previous derived sum rules still hold.
This includes e.g.\ an effectively one-dimensional system in planar geometry that evolves to the same bulk state at the left and right boundaries or if the boundary conditions are periodic. 
In both cases left and right boundary terms are equal up to a minus sign and hence cancel each other. 
This example can be generalized straightforwardly to more complex geometries.

For non-vanishing net boundary terms additional contributions arise in the above sum rules. 
These contributions occur if the system develops different (bulk) states, e.g.\ for $x\to\pm \infty$ as is relevant for bulk phase separation (see  the section below).
We demonstrate that such cases can be systematically treated in the
current framework, by exemplary considering the total internal
force.
 Then boundary force contributions arise due to an imbalance of
“outside” particles that exert forces on “inside” particles. The
outside particles are per definition excluded from the accounting of
the total internal force exerted by all particles inside of~$V$. 
 The sum of all interactions between inside particles vanishes due to the global internal Noether sum rule \eqref{EQFint}.  
 For
simplicity we restrict ourselves to systems that interact via
short-ranged pairwise central forces, where $\textbf{F}_{ij}$
indicates the force on particle $i$ exerted by particle $j$.
 So only forces exerted from an inside to an outside particle contribute.
  The total internal force that acts on $V$ is hence $\textbf{F}_\text{int}^\text{tot}
= \langle \sum_{ij}' \textbf{F}_{ij} \rangle$, where the restricted
sum (prime) runs only over those $i \in V$ and $j \in \bar{V}$, where
$\bar V$ indicates the complement of $V$. 
 The total internal force between particles inside of $V$, i.e.\ $i \in V$ and $j \in V$, vanishes due to \eqref{EQFint}.
We then rewrite
$\textbf{F}_\text{int}^\text{tot}$ via inserting the identity $\int
d\rv\delta(\rv)=1$ twice into the average. Then the restrictions of
the sums can be transferred to restrictions on the spatial integration
domains. As a result the total internal force acting on~$V$ can be
expressed via correlation functions as
\begin{align}
  \Fv_\text{int}^\text{tot} 
  & =\int_V d \rv \int_{\bar{V}} d \rv' 
  \Big\langle
  \sum_{i,j \neq i} \delta(\rv-\rv_i) \delta(\rv'-\rv_j)
  \textbf{F}_{ij} \Big\rangle
  \\
  &= -\int_V d \rv \int_{\bar{V}} d \rv' 
  \rho(\rv) \rho(\rv') g(\rv,\rv')
  \nabla\phi(|\rv-\rv'|),
  \label{EQFintTotPairForm}
\end{align}
where $\phi(r)$ indicates the interparticle pair potential as a
function of interparticle distance $r$. In order to obtain the form
\eqref{EQFintTotPairForm} we have identified the many-body definition
of the radial distribution function $g(\rv,\rv')= \big\langle
\sum_{i,j\neq i} \delta(\rv-\rv_i) \delta(\rv'-\rv_j)
\big\rangle/(\rho(\rv)\rho(\rv'))$. Recall that the pair distribution
function $g$ and the density-density correlation function $H_2$, as
used in \eqref{EQBobH2}-\eqref{EQHn}, 
are related via
$H_2(\rv,\rv')=(g(\rv,\rv')-1)\rho(\rv)\rho(\rv')+\delta(\rv-\rv')\rho(\rv)$.
Equation \eqref{EQFintTotPairForm} still holds for non-conservative
interparticle forces, when $-\nabla\phi(|\rv-\rv'|)$ is replaced by
the (nongradient) interparticle force field.
 We demonstrate in the following section the practical relevance of these considerations.

\begin{figure}
  \includegraphics[width=0.79\columnwidth,angle=0]{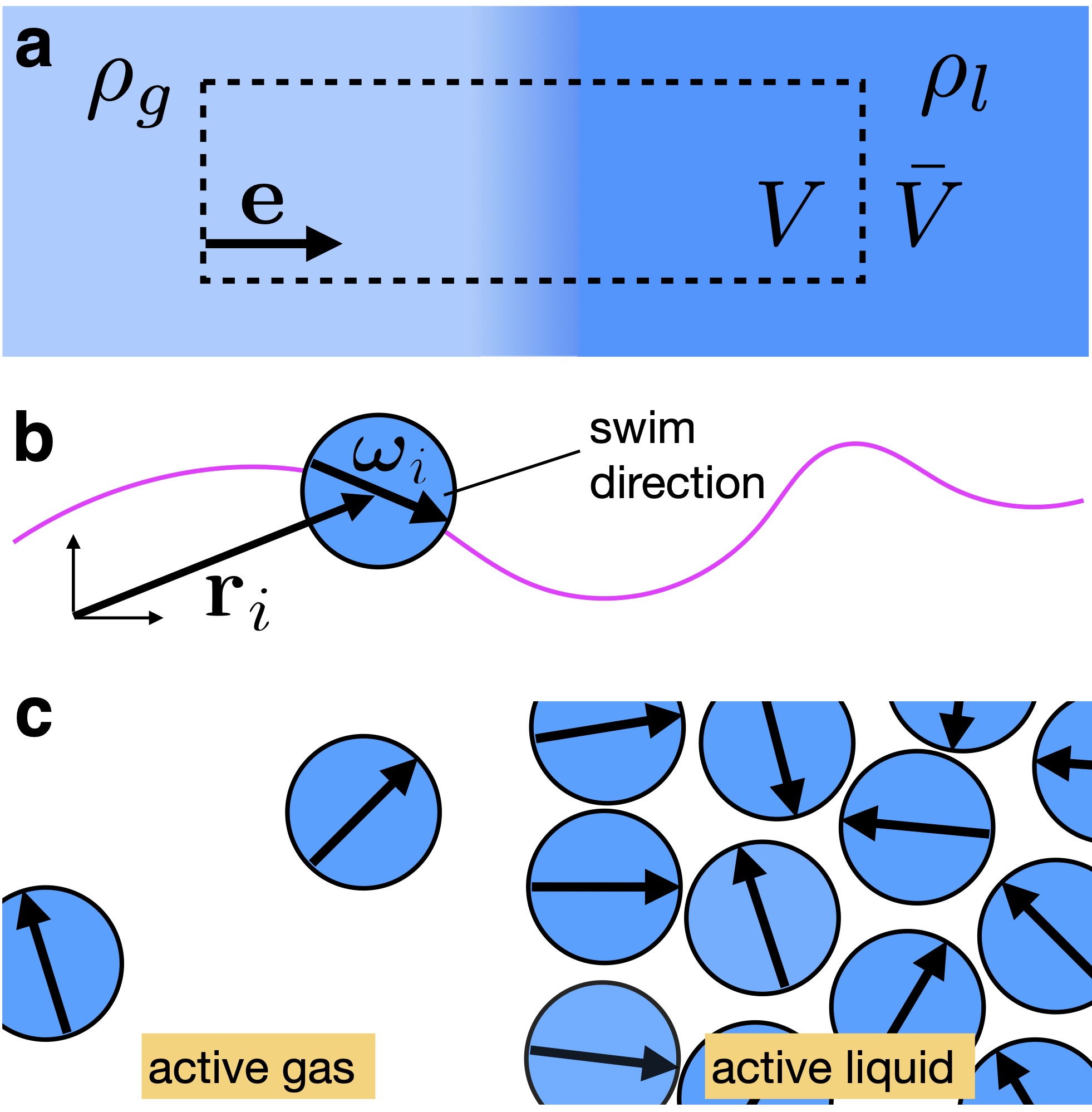}
  \caption{ \textbf{Phase separation into macroscopically distinct phases.} \textbf{a}~Illustration of the geometry of phase separation of passive or active particles. Shown are the gaseous $\rho_g$ and liquid $\rho_l$ plateau values of the
    density profile (indicated by the colour gradient) and the direction vector $\bf e$ normal to the interface. The total system volume  $V+\bar V$ consists of the subvolume $V$ and its complement $\bar V$.
    \textbf{b}~Illustration of 
an active Brownian particle (blue disc) with position $\textbf{r}_i$ and orientation $\boldsymbol{\omega}_i$ undergoing translational and rotational diffusion. The self propulsion along $\boldsymbol{\omega}_i$ creates directed motion as indicated by the trajectory (magenta line).
\textbf{c}~Schematics of motility-induced phase separation into active gas (left) and active liquid phases (right). The arrows indicate the orientations $\boldsymbol{\omega}_i$ of the active particles. The interface is polarized. 
}
  \label{fig4}
\end{figure}

\mysubsec{Phase coexistence}{SECphaseCoexistence}We turn to situations of
phase coexistence. As we demonstrate, considering a large, but finite
subvolume~$V$ of the entire system is useful  
but it also requires to take boundary terms into account.  
Here we take $V$ to be
cuboidal and to contain the free (planar) interface between two
coexisting phases, see Fig.~\ref{fig4}(a) for a graphical
illustration. The volume boundaries parallel to the interface are
taken to be seated deep inside either bulk phase.  The internal force
contributions on those faces of $V$ that ``cut through'' the
interface, i.e.\ have a normal that is perpendicular to the interface
normal, vanish by symmetry. It remains to evaluate 
\eqref{EQFintTotPairForm} over each of the two faces in the respective
bulk region. Therefore the position dependences simplify to $\rho(\textbf{r}) =
\rho_b= \text{const}$, where $\rho_b$ indicates the bulk number
density, and the inhomogeneous pair distribution function simplifies
as $g(\textbf{r},\textbf{r}') = g(|\textbf{r}-\textbf{r}'|)$.
Furthermore only force contributions collinear with $\ev$, the outer
interface normal of the considered bulk phase $b$, contribute. For a
single face in bulk phase $b$, it is straightforward to show that the
result is the virial pressure multiplied by the interface area $A$,
i.e.\ the force $ A p^b_\text{int} \textbf{e}$, where the internal
interaction pressure $p^b_\text{int}$ in phase $b$ is e.g.\ given via
the Clausius virial~\cite{hansen2013}, 
 $p^b_\text{int}= - \frac{\pi}{2} \rho^2_b \int\limits_0^{r_0} d r \, r^2 g(r) \frac{d \phi}{d r}$ in two spatial dimensions. (The argument remains general.) The constant $r_0$ denotes the range of the interparticle interactions.

The total force density balance for equilibrium phase separation contains thermal diffusion and the internal force density, which cancel each other,
\begin{align}
0=- k_\text{B} T \nabla \rho(\textbf{r}) + \rho(\textbf{r}) \textbf{f}_\text{int}(\textbf{r}).
\end{align}
Integration over the volume $V$ yields the total force which is proportional to the pressure.
Hence the total internal force $\textbf{F}^\text{tot}_\text{int} = \int_V d \textbf{r} \, \rho \textbf{f}_\text{int}$ on $V$, cf.~\eqref{EQFintTotPairForm}, amounts
to the pressure difference $(p_{\rm int}^g-p_{\rm int}^l)A \ev$, where $\ev$ is the (unit vector) normal of the interface, pointing from,
say, the gas (index $g$) to the liquid phase (index $l$).
 The total
diffusive force is $(p_{\rm id}^g-p_{\rm id}^l)A\ev$ with the pressure
of the ideal gas $p_{\rm id}=k_\text{B}T\rho$, evaluated at the gas
($\rho_g$) and liquid bulk density ($\rho_l$). As there are no
external forces ($V_{\rm ext}\equiv 0$) then requesting the volume $V$
to be forcefree amounts to $p_{\rm tot}^g=p_{\rm tot}^l$, where the
total pressure is the sum $p_{\rm tot}=p_{\rm id}+p_{\rm int}$. Hence
the boundary consideration yields the mechanical equilibrium condition
of equality of pressure in the coexisting phases. 

We conclude that the internal interactions that occur across the free
interface do not influence the (bulk) balance of the pressure at phase
coexistence, as the net effect of these interactions vanishes. At the
heart of this argument lies Noether's theorem for invariance against
spatial displacements.

\mysubsec{Anisotropic particles}{SECanisotropicParticles} 
We turn to
anisotropic interparticle interactions, where $\ov_i$, $\boldsymbol{\varpi}_i$ are two perpendicular unit vectors that describes the
particle orientation in space. Such systems are described by an
interparticle interaction potential $u(\rv^N, \ov^N, \boldsymbol{\varpi}^N)$, which is
assumed a priori to be invariant under spatial translations.  Similarly one-body fields in general depend on position $\rv$ and orientations $\ov$ and $\boldsymbol{\varpi}$ of the particles, e.g.\ $V_\text{ext}(\rv,\ov,\boldsymbol{\varpi})$ for the external field. The fully resolved one-body density
distribution is
$\rho(\rv,\ov,\boldsymbol{\varpi},t)=\langle\sum_i\delta(\rv-\rv_i)\delta(\ov-\ov_i)\delta(\boldsymbol{\varpi}-\boldsymbol{\varpi}_i)\rangle$.

It is straightforward to ascertain that all the above (force) sum
rules for translation remain valid, as the orientations are unaffected
by translations, upon trivially generalizing from position-only to
position-orientation integration, $\int d\rv\to \int d\rv d\ov d\boldsymbol{\varpi}$  etc.

In the following for simplicity of notation we first consider uniaxial particles. Uniaxial particles depend only on one single orientation $\ov_i$, as the particles are rotationally invariant around this vector.
Hence the $\boldsymbol{\varpi}$-dependence of both the one- and many-body quantities vanish and the total integral simplifies to $\int d\rv d\ov $.

\mysubsec{Motility induced phase separation}{SECMIPS}
We use active Brownian particles
 as an example for uniaxial particles. For simplicity we consider spherical particles (discs) in two dimensions. 
The particles repel each other and they undergo self-propelled motion along their orientation vector $\ov$. Hence an additional one-body force $\gamma s \ov$ acts on each swimmer, with  $\gamma$ the friction constant and $s$ the speed of free swimming. 
This self-propulsion creates characteristic trajectories (see Fig. \ref{fig4}(b) for a schematic) which are also affected by thermal diffusion (omitted in the schematic) of the particle position $\textbf{r}_i$ and orientation $\ov_i$.
Experimental realizations of active Brownian particles include e.g.\ Janus colloids driven by photon nudging \cite{nudge1,nudge2,nudge3,nudge4,nudge5}.
If the density is high enough, motility induced phase separation (MIPS) into an active gas and active liquid phase occurs for high enough values of the swim speed $s$, cf. Fig. \ref{fig4}(c).

The force density balance (see e.g.\ the work of Hermann \textit{et al.} \cite{hermann2019pre}) of such a system in steady state (no time dependence) is
\begin{align}
\gamma \textbf{J}(\textbf{r},\ov) = &- k_\text{B} T \nabla \rho(\textbf{r},\ov) + \rho(\textbf{r},\ov) \textbf{f}_\text{int}(\textbf{r},\ov) \nonumber\\
 &+ \gamma s \rho(\textbf{r},\ov) \ov + \rho(\textbf{r},\ov) \textbf{f}_\text{ext}(\textbf{r},\ov). \label{EQAphase}
\end{align}
The (negative) frictional force density on the left hand side is balanced with the ideal gas contribution (first term), the interparticle interactions (second term), the self-propulsion (third term) and the external contribution (fourth term) on the right hand side.
As in case of the equilibrium phase separation we integrate \eqref{EQAphase} over the volume $V$ (see Fig. \ref{fig4}(a) for an illustration) and over all orientations $\ov$. In the following we discuss each term separately. For simplicity we assume planar geometry of the system and we assume a vanishing external force, $\textbf{f}_\text{ext}(\textbf{r},\ov)=0$.
Here, the interaction contribution $p_{\rm int}A\ev$ is obtained as above in equilibrium via \eqref{EQFintTotPairForm}, but the
virial is averaged over the nonequilibrium steady state many-body
probability distribution. 
The integral over the current  $\int d\textbf{r} d\ov \textbf{J}$ 
is assumed to vanish  in steady state.

The total swim force that acts on $V$
contributes to the total force. In the considered situation, the swim
force is entirely due to the polarization $\textbf{M}_\text{tot} = \int d \textbf{r} d \ov \, \ov \rho$ of the free interface in MIPS. Particles at
the interface tend to align against the dense phase \cite{paliwal2018} if
they interact purely repulsively (cf. Fig. \ref{fig4}(c)), and they align against the dilute phase if
interparticle attraction is present \cite{paliwal2017activeLJ}.  No
such spontaneous polarization occurs in bulk. The interface
polarization is a state function of the coexisting phases
\cite{hermann2020polarization} as verified both experimentally \cite{leipzig1} and numerically \cite{leipzig2}.
The total swim force that acts on $V$ is
$(p_{\rm swim}^g-p_{\rm swim}^l)A \ev$, where $p_{\rm swim}^b=\gamma s J_b/(2D_{\rm rot})$, with $J_b$ the bulk current in the forward
direction $\ov$ and $D_{\rm rot}$ indicating the rotational diffusion
constant. Apart from the ideal term no further forces act, cf.\ the force density balance \eqref{EQAphase}. The integral over the ideal term is similar to the total equilibrium ideal contribution, $(p^g_\text{id}-p^l_\text{id}) A \ev$.
Combination of all results from integration yields the total force balance. 

As the negative integral over the force density defines the nonequilibrium pressure, the volume $V$ being force free amounts to $p_{\rm tot}^g=p_{\rm tot}^l$, where
$p_{\rm tot}=p_{\rm id}+p_{\rm int}+p_{\rm swim}$. 
Hermann \textit{et al.} \cite{hermann2019pre} demonstrates the splitting of $p_{\rm int}$ into
adiabatic and superadiabatic contributions and presents results for
the phase diagram based on approximate forms for the interparticle
interaction contributions. 
The pressure, especially its swim contribution is 
defined in various different ways in the literature  \cite{pressure1,pressure2,pressure3,pressure4,pressure5,pressure6}.

We conclude that the internal interactions
that occur across the free interface do not influence the (bulk)
balance of the pressure at phase coexistence, as the net effect of
these interactions vanishes. In essence this argument follows from 
Noether's theorem for invariance against spatial displacements.

\begin{figure}
  \includegraphics[width=0.59\columnwidth,angle=0]{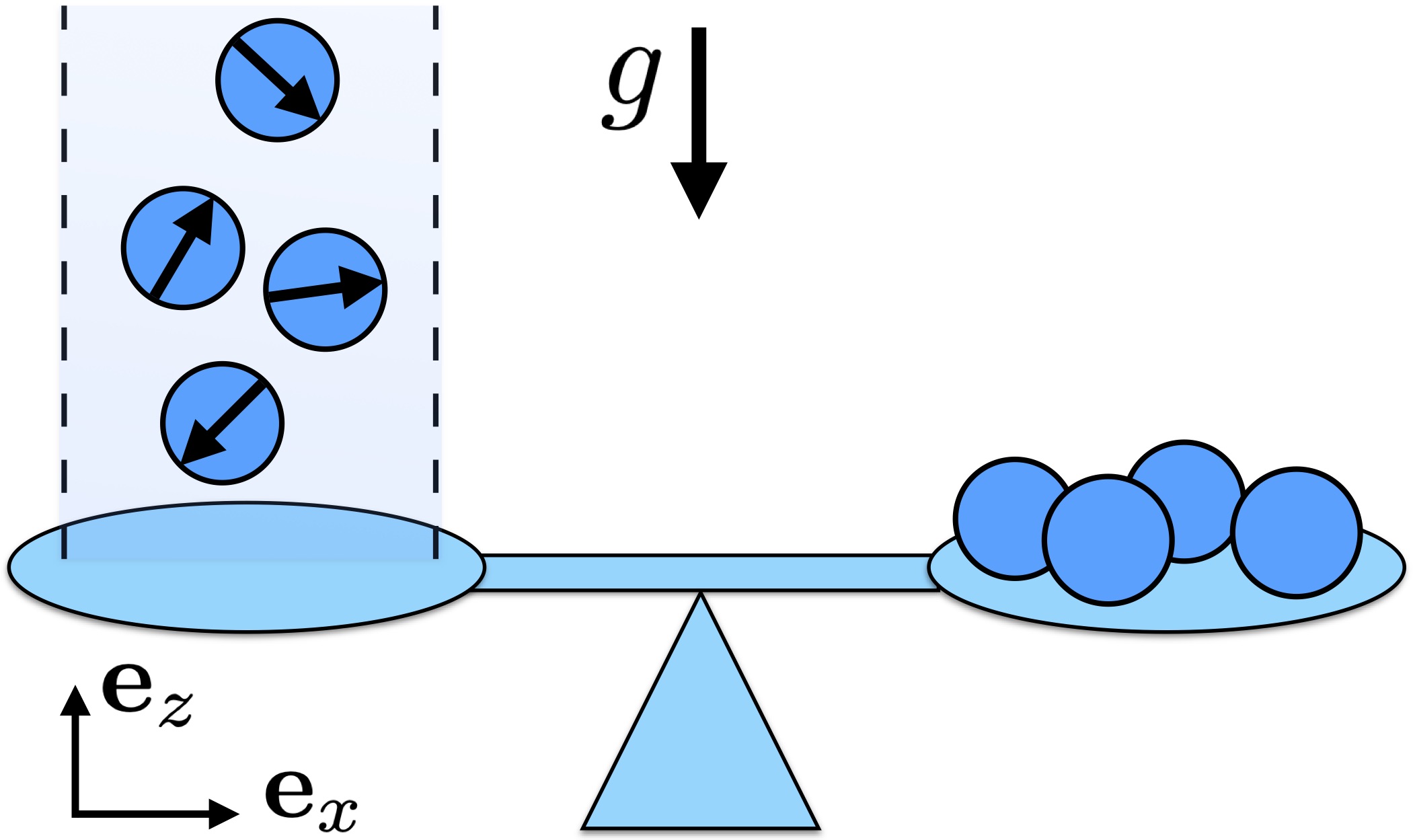}
  \caption{  \textbf{Illustration of sedimentation of active Brownian particles under gravity $g$.} 
  The active particles with orientation $\boldsymbol{\omega}$ (black arrows) are confined by a lower wall and periodic boundary conditions on the sides (dashed lines).  
  The total force that the swimming particles exert on the bottom wall (left scale pan) is equal to their weight (right scale pan) in steady states of the system.
    }
  \label{fig5}
\end{figure}

\mysubsec{Active sedimentation}{SECSedi}
Sedimentation under the influence of gravity is a ubiquitous phenomenon in soft matter that has attracted considerable interest, e.g.\ for colloidal mixtures \cite{dani1,dani2,dani3,dani4} and for active systems \cite{sedi1,sedi2,sedi3,sedi4}.
As sedimentation is a force driven phenomenon the Noether sum rules apply directly, as we show in the following.

We assume that the system is translationally invariant in the $x$-direction and that an impenetrable wall at $z=0$ acts as a lower boundary of the system \cite{lee2013} (cf.\ Fig. \ref{fig5}). We assume the wall-particle interaction potential to be short-ranged. Its precise form is irrelevant for the following considerations. 
The force density balance for such a system is \eqref{EQAphase} with the external force field chosen as
\begin{align}
\textbf{f}_\text{ext}(\textbf{r},\ov) = - m g \textbf{e}_z + \textbf{f}_\text{wall}(\textbf{r},\ov), \label{EQSedi}
\end{align}
where $m$ denotes the mass of a particle, $g$ is the gravitational acceleration and $\textbf{e}_z$ indicates the unit vector in $z$-direction. Hence the external force field $\textbf{f}_\text{ext}$ consists of gravity and the wall contribution $\textbf{f}_\text{wall}$.

To proceed we integrate the force density balance over all positions $\textbf{r}$ in the volume $V$ and over all orientations $\ov$. 
The total integral of the density distribution (per radiant) $\int d\textbf{r} d\ov  \rho$, as appears in the gravitational term, gives the total number of particles $N$. 
The integral over the total current $\int d\textbf{r} d\ov \, \textbf{J}$ is proportional to the center of mass velocity 
$\textbf{v}_\text{cm}(t) = \int d\textbf{r} d\ov \, \rho \textbf{v} / \int d\textbf{r} d\ov \,\rho = \frac{1}{N} \int d\textbf{r} d\ov \, \textbf{J}$. Here $\textbf{v}_\text{cm}(t)$ is a global quantity and hence it is independent of both position and orientation. 
We first only consider steady states,  so the center of mass velocity $\textbf{v}_\text{cm}(t)=0$ and hence the total current vanishes. 
The total thermal diffusion term vanishes because $\int_V d\textbf{r}\, \nabla \rho = \int_{\delta V} d \textbf{S}\, \rho = 0$ as there is no contribution of $\rho$  from the boundaries $\partial V$ of the integration volume $V$. At the upper and lower boundary the density is zero as it vanishes in the wall and also for $z \to \infty $. The left and the right boundary contributions cancel each other as
the density is independent of $x$ due to translational invariance.
The total internal interaction force density vanishes, $\textbf{F}_\text{int}^\text{tot} = \int d\textbf{r} d\ov \, \rho \textbf{f}_\text{int} =0$, using the global internal Noether sum rule \eqref{EQFint}. 
The integrated swim force density is proportional to the total polarization $\textbf{M}_\text{tot}= \int d\textbf{r} d\ov \, \ov \rho $. This quantity vanishes, $\textbf{M}_\text{tot} =0$, as there is no net flux through the boundaries in steady state (see Eq. (10) by Hermann \textit{et al.} \cite{hermann2020polarization}). 
Combination of all integrals yields the relation
\begin{align}
\textbf{F}^\text{tot}_\text{wall} = \int d\textbf{r} d\ov \, \rho \textbf{f}_\text{wall}  = m g N \textbf{e}_z.  \label{EQFish}
\end{align}
Hence the $z$-component of the total force on the wall $F^\text{tot}_\text{wall}$ is equal to the total gravitational force acting on all particles (see Fig. \ref{fig5} for a graphical representation). Equation \eqref{EQFish} of course also holds for passive colloids ($s=0$).

Keeping the translational invariance in the $x$-direction, we next turn to time dependent systems. Therefore all one-body field in \eqref{EQAphase} additionally depend on the time $t$. 
Integration of the  force density balance \eqref{EQAphase} gives identical results for the thermal diffusion, the internal force density and for the  gravitational contribution as in the above case of steady state. Even for the time-dependent dynamics these integrals are independent of time $t$. 
The total wall force density is given as $\textbf{F}_\text{wall,tot}=F^\text{tot}_\text{wall} \textbf{e}_z$ and it only acts 
along the unit vector in the $z$-direction $ \textbf{e}_z$, due to the symmetry of the system.
The integral of the self-propulsion term is still proportional to  the total polarisation. 
However, the total polarization does not vanish in general but decays exponentially (see Eq. (21) by Hermann \textit{et al.} \cite{hermann2020polarization}),
\begin{align}
\textbf{M}_\text{tot}(t) = \textbf{M}_\text{tot}(0) e^{-D_\text{rot}t},
\end{align} 
where $\textbf{M}_\text{tot}(0)$ indicates the initial polarization at time $t=0$ and the time constant $1/D_\text{rot}$ is the inverse rotational diffusion constant.
Similarly, integration of the current still gives the (time-dependent) center of mass velocity $\textbf{v}_\text{cm}(t)$. 

Insertion of these results into the spatial and orientational integration of  \eqref{EQAphase} leads to the total friction force
\begin{align}
\gamma N \textbf{v}_\text{cm}(t) = s \gamma \textbf{M}_\text{tot}(0)\, e^{-D_\text{rot} t} + F^\text{tot}_\text{wall}(t)\, \textbf{e}_z- mgN \textbf{e}_z, \label{EQvcm}
\end{align}
which is hence a direct consequence of the Noether sum rule \eqref{EQFint}.
We find that the $x$-component of the center of mass velocity decays simultaneously with the total polarization, cf.\ the first term on the right hand side of \eqref{EQvcm}. The $z$-component of $\textbf{v}_\text{cm}(t)$ depends on $\textbf{M}_\text{tot}(t)$ and additionally on the time-dependent total force exerted by the wall and the total graviational force.
Hence measuring the total force on the wall (i.e.\ by weighing, cf. Fig. \ref{fig5}) and knowledge of the total initial polarization and the total particle number allows  one to determine the center of mass velocity.
Note that in the limit of long times, $t\to\infty$,  the total polarization vanishes and this system evolves to a steady state. Hence
the center of mass velocity vanishes and \eqref{EQFish} is recovered.
As we have demonstrated both statements \eqref{EQFish} and \eqref{EQvcm} ultimately follow from the global Noether identity \eqref{EQFint}.

\mysubsec{Rotational invariance}{SECrotationalInvariance}
We return to the  
general case and initially consider spatial rotations in systems of
spheres, i.e.\ systems where $u(\rv^N)$ depends solely on (relative)
particle positions, and where it is invariant under global rotation of
all $\rv^N$ around the origin. We parameterize the rotation by a
vector $\nv$.  The direction of $\nv$ indicates the rotation axis and
the modulus $|\nv|$ is the angle of rotation. To lowest nonvanishing
order, the rotation amounts to $\rv\to\rv+\nv\times\rv$.  One-body
functions change accordingly: the external potential undergoes $V_{\rm
  ext}(\rv)\to V_{\rm ext}(\rv)+\delta V_{\rm ext}(\rv)$, with $\delta
V_{\rm ext}(\rv)=(\nv\times\rv)\cdot\nabla V_{\rm ext}(\rv)$ and the
density profile $\rho(\rv)\to\rho(\rv)+\delta\rho(\rv)$ with
$\delta\rho(\rv)=(\nv\times\rv)\cdot\nabla\rho(\rv)$. Much of the
reasoning of the above case of spatial displacement can be applied
readily: $\Omega[V_{\rm ext}]$ is invariant under the rotation, and
$\delta\Omega=\int d\rv(\delta\Omega/\delta V_{\rm ext}(\rv))\delta
V_{\rm ext}(\rv)=\int d\rv\rho(\rv)(\nv\times\rv)\cdot\nabla V_{\rm
  ext}(\rv)=0$. As the rotation vector $\nv$ is arbitrary, we can
conclude that the total external torque $\Torque_{\rm ext}^{\rm tot}$
vanishes in equilibrium \cite{baus1984},
\begin{align}
  \Torque_{\rm ext}^{\rm tot}
  \equiv-\int d\rv \rho(\rv)(\rv\times\nabla V_{\rm ext}(\rv))=0.
  \label{EQTorqueExtTotVanishes}
\end{align}
As this holds true for any form of the applied $V_{\rm ext}(\rv)$, we
can differentiate with respect to $V_{\rm ext}(\rv')$, and obtain
\cite{baus1984}
\begin{align}
  &\rv\times\nabla\rho(\rv) =
  -\int d\rv' \beta H_2(\rv,\rv') (\rv'\times\nabla' V_{\rm ext}(\rv')),
  \label{EQtorqueH2}
  \\
  &  \sum_{\alpha=1}^n(\rv_\alpha\times\nabla_\alpha H_n)= \notag\\&  
  \quad -
  \int d\rv_{n+1}\beta V_{\rm ext}(\rv_{n+1})
  (\rv_{n+1}\times\nabla_{n+1} H_{n+1}).
  \label{EQtorqueHn}
\end{align}
The excess free energy density functional $F_{\rm exc}[\rho]$ can be
treated accordingly. It is invariant under rotation, as its sole
dependence is on $u(\rv^N)$, which by assumption is rotationally
invariant. Analogous to this reasoning, we obtain the result that
the total interparticle adiabatic torque $\Torque_{\rm ad}^{\rm tot}$
vanishes,
\begin{align}
  \Torque_{\rm ad}^{\rm tot}=\int
  d\rv\rho(\rv)(\rv\times\fv_{\rm ad}(\rv))=0.
  \label{EQTorqueAdiabaticTotVanishes}
\end{align}
Differentiation with respect to the independent field $\rho(\rv)$ once
and $n$ times yields the respective identities
\cite{baus1984,lovett1991}:
\begin{align}
  \rv\times\nabla c_1(\rv) &= \int d\rv' c_2(\rv,\rv')
  (\rv'\times\nabla'\rho(\rv')),
  \label{EQtorquec2}\\
  \sum_{\alpha=1}^n (\rv_\alpha\times\nabla_\alpha c_n) &=
  -\int d\rv_{n+1} \rho(\rv_{n+1})(\rv_{n+1}\times\nabla_{n+1} c_{n+1}).
  \label{EQtorquecn}
\end{align}

The multi-body versions of the theorems of vanishing total external
\eqref{EQTorqueExtTotVanishes} and adiabatic internal
\eqref{EQTorqueAdiabaticTotVanishes} torques are, respectively, 
\begin{align}
\int
d \rv_1 V_\text{ext}(\rv_1) \dotsc \int d \rv_n V_\text{ext}(\rv_n)
(\rv_\alpha \times \nabla_\alpha H_n)=0, \label{EQHnrot}
\end{align}
 and 
\begin{align}
\int d \rv_1 \rho(\rv_1)
\dotsc \int d \rv_n \rho(\rv_n) (\rv_\alpha \times
\boldsymbol{\nabla}_\alpha c_n)=0, \label{EQcnrot}
\end{align}
 for $\alpha = 1\dotsc n$.
These identities are respectively derived from \eqref{EQtorqueH2} by
multiplying with $V_\text{ext}(\rv)$, integrating over $\textbf{r}$
and exploiting \eqref{EQTorqueExtTotVanishes}, and from
\eqref{EQtorquec2} by multiplying with $\rho(\rv)$, integrating over
$\textbf{r}$ and exploiting \eqref{EQTorqueAdiabaticTotVanishes}, and
iteratively repeating for each order~$n$.

On the many-body level, it is straightforward to see that the total
internal torque $-\sum_i (\rv_i\times\nabla_i u(\rv^N))=0$. Hence, as
this identity holds for each microstate, its general, nonequilibrium
average vanishes, $\Torque_{\rm int}^{\rm tot}=0$. The force field
splitting $\fv_{\rm int}=\fv_{\rm ad}+\fv_{\rm sup}$ induces
corresponding additive structure for the internal total torque:
$\Torque_{\rm int}^{\rm tot}=\Torque_{\rm ad}^{\rm tot}+\Torque_{\rm
  sup}^{\rm tot}$, with the total superadiabatic (internal) torque
$\Torque_{\rm sup}^{\rm tot}=\int d\rv\rho(\rv)[\rv\times\fv_{\rm
    sup}(\rv,t)]$. As $\Torque_{\rm int}^{\rm tot}=\Torque_{\rm
  ad}^{\rm tot}=0$, we conclude $\Torque_{\rm sup}^{\rm tot}=0$,
$\forall t$.

To apply Noether's Theorem to the power functional, we consider an
instantaneous rotation, with infinitesimal angular velocity $\dot\nv$
at time $t$. The effect is a change in current $\Jv\to\Jv+\delta \Jv$
with $\delta\Jv=(\dot\nv\times\rv)\rho(\rv)$. Correspondingly, the
velocity field acquires an instantaneous global rotational
contribution, according to
$\vel(\rv,t)\to\vel(\rv,t)+\dot\nv\times\rv$.  The superadiabatic
excess power functional is invariant under this operation and hence
$P_t^{\rm exc}[\rho,\Jv]=P_t^{\rm exc}[\rho,\Jv+\delta\Jv]=P_t^{\rm
  exc}[\rho,\Jv]+\int d\rv (\delta P_t^{\rm
  exc}[\rho,\Jv]/\delta\Jv(\rv,t))\cdot(\dot\nv\times\rv)\rho(\rv,t)$. As
$\dot\nv$ is arbitrary, we can conclude
\begin{align}
  \Torque_{\rm sup}^{\rm tot} &= \int d\rv
  \rho(\rv,t)(\rv\times\fv_{\rm sup}(\rv,t))=0,
  \quad \forall t,
  \label{EQTorqueSupTotalVanishes}
\end{align}
as is consistent with the result of the above many-body derivation.
As \eqref{EQTorqueSupTotalVanishes} holds for any (trial)
$\Jv(\rv,t)$, the derivative of \eqref{EQTorqueSupTotalVanishes} with
respect to $\Jv(\rv,t)$ vanishes. Hence
\begin{align}
  \int d\rv \rho(\rv,t) (\rv\times {\sf M}_2(\rv,\rv',t)) = 0,
  \label{EQsupTorqueM2}
\end{align}
where the cross product with a tensor is defined via contraction with
the Levi-Civita tensor.  At $n$-th order we obtain 
\begin{align}
\int
d\rv_n\rho(\rv_n,t)(\rv_n\times{\sf M}_n)=0.
\end{align}
 This identity and
\eqref{EQsupTorqueM2} express the vanishing of the total
superadiabatic torque, when resolved on the $n$-body level of (time
direct) correlation functions.

\begin{figure}
  \includegraphics[width=0.59\columnwidth,angle=0]{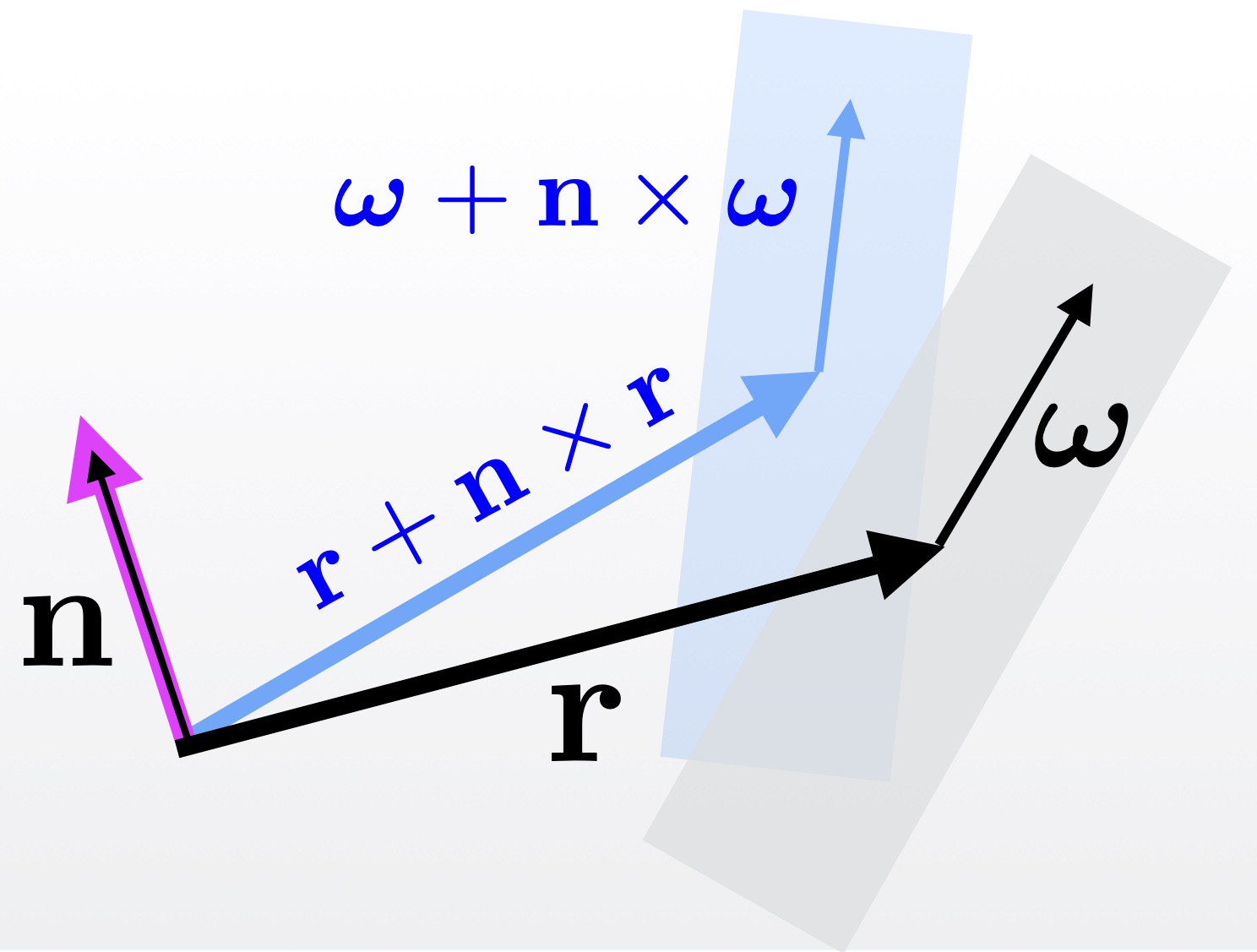}
  \caption{ \textbf{Illustration of the rotation
    operation of uniaxial particles.} 
     The particles (rectangular shapes) at position
    $\rv$ and with orientation $\ov$  are shown in original (black) and rotated
    (blue) configuration, where $\nv$ indicates the rotation axis
    (direction) and angle (length). 
    }
  \label{fig6}
\end{figure}

\mysubsec{Orbital and spin coupling}{SECSpin}
The case of rotational symmetry of uniaxial particles is clearly more
complex, as both particle coordinates and particle orientations are
affected by a global (``rigid'') operation on the entire
system, i.e.\ both positions and orientations are rotated
consistently. Noether's Theorem ensures though that this operation is indeed
the fundamental one, and that the physically expected coupling of
orbital and spinning effects will naturally and systematically
emerge. Here we use (common) terminology for referring to spin as
orientation vector rotation (i.e. particle rotation around its
center), as opposed to orbital rotation (of position vector) around
the origin of position space. Hence all the above considered torques
that already occur in systems of spheres are of orbital nature. These
of course remain relevant for anisotropic particles, but the
nontrivial orientational behaviour of the latter will generate
additional spin torques. The global rotation consists of an orbital
part, $\rv\to\rv+\nv\times\rv$, and a spin part,
$\ov\to\ov+\nv\times\ov$; see Fig.~\ref{fig6} for a graphical
representation.

 For anisotropic systems the external field naturally acquires
dependence on position $\rv$ and orientation $\ov$, i.e.\ $V_{\rm ext}(\rv,\ov)$, where
$-\nabla V_{\rm ext}(\rv,\ov)$ is the external force field as before,
 and $-\ov\times\nabla^\omega V_{\rm ext}(\rv,\ov)$ is the external
torque field, where $\nabla^\omega$ is the derivative with respect to
$\ov$ in orientation space.
 Hence the induced change of external potential is
$\delta V_{\rm ext}(\rv,\ov)=(\nv\times\rv)\cdot\nabla V_{\rm
  ext}+(\nv\times\ov)\cdot\nabla^\omega V_{\rm ext}$. This change
leaves the grand potential $\Omega[V_{\rm ext}]$ invariant, hence
$\delta\Omega=\int d\rv d\ov (\delta\Omega/\delta V_{\rm
  ext}(\rv,\ov))\nv\cdot (\rv\times\nabla V_{\rm
  ext}+\ov\times\nabla^\omega V_{\rm ext}) =0$, from which we identify
the rotational Noether Theorem for the total external torque of
anisotropic particles:
\begin{align}
  \Torque_{\rm ext}^{\rm tot} = -\int d\rv d\ov \rho(\rv,\ov)
  (\rv\times \nabla V_{\rm ext} + 
  \ov\times\nabla^\omega V_{\rm ext}) &= 0.
  \label{EQTarazonaEvans1global}
\end{align}
Differentiation with respect to $V_{\rm ext}(\rv,\ov)$ yields
\begin{align}
  &\rv\times\nabla\rho + \ov\times\nabla^\omega\rho =
  \label{EQTarazonaEvans1}
  \\&
  - \negthinspace \int \negthinspace \negthinspace d\rv' d\ov'
  \beta H_2(\rv,\ov,\rv',\ov')
  (\rv'\times\nabla' V_{\rm ext}'+\ov'\times\nabla'^\omega V_{\rm ext}'),
  \notag\\
  & \sum_{\alpha=1}^n(\rv_\alpha\times\nabla_\alpha H_n
  +\ov_\alpha\times\nabla^\omega_\alpha H_n) =\notag\\
  & -\int d \rv' d\ov' \beta V_{\rm ext}'
  (\rv'\times\nabla' H_{n+1}
  +\ov'\times\nabla'^\omega H_{n+1}),
  \label{EQTarazonaEvans1hierarchy}
\end{align}
where we have used $V_{\rm ext}'=V_{\rm ext}(\rv',\ov')$ as a
shorthand, $H_2(\rv,\ov,\rv',\ov')=-\delta \rho(\rv,\ov)/\delta \beta
V_{\rm ext}(\rv',\ov')$ is the density-density correlation function,
and its $n$-body version $H_n=H_n(\rv_1,\ov_1\ldots\rv_n,\ov_n)$ with
$H_{n+1}=\delta H_n/\delta \beta V_{\rm ext}(\rv_{n+1},\ov_{n+1})$; in
\eqref{EQTarazonaEvans1hierarchy} the prime refers to the $n+1$th
degrees of freedom.  Crucially, on all levels of $n$-body correlation
functions, the spin and orbital torques remain coupled.

Turning to internal torques, the change of density upon global
rotation is $\delta\rho(\rv,\ov)=\nv\cdot(\rv\times\nabla\rho+
\ov\times\nabla^\omega\rho)$ and the net effect on the excess free
energy is $\delta F_{\rm exc} = \int d\rv d\ov(\delta F_{\rm
  exc}[\rho]/\delta\rho(\rv,\ov))\nv\cdot
(\rv\times\nabla\rho+\ov\times\nabla^\omega\rho)=0$. We hence obtain
\begin{align}
  \Torque_{\rm ad}^{\rm tot} &=
  \int d\rv d\ov \rho(\rv,\ov)
  (\rv\times \fv_{\rm ad} + \torque_{\rm ad})
  =0,
  \label{EQTarazonaEvans2global}
\end{align}
where the adiabatic spin torque field is $\torque_{\rm
  ad}(\rv,\ov)=-\ov\times\nabla^\omega \delta F_{\rm exc}[\rho]/\delta
\rho(\rv,\ov)$. From differentiation with respect to $\rho(\rv,\ov)$
we obtain
\begin{align}
  &  \rv\times\fv_{\rm ad} + \torque_{\rm ad} =\notag\\
  &  \int d\rv' d\ov' c_2(\rv,\ov,\rv',\ov')
  (\rv'\times\nabla'\rho'+\ov'\times\nabla'^\omega\rho'),
  \label{EQTarazonaEvans2}\\
  &  \sum_{\alpha=1}^n(\rv_\alpha\times\nabla_\alpha c_n
  +\ov_\alpha\times\nabla_\alpha^\omega c_n) = \notag\\
  &  -\int d\rv'd\ov' \rho'(\rv'\times\nabla' c_{n+1}
  +\ov'\times\nabla'^\omega c_{n+1}),
  \label{EQTarazonaEvans2hierarchy}
\end{align}
where $\rho'=\rho(\rv',\ov')$.  Multi-body versions of
\eqref{EQTarazonaEvans1global} and \eqref{EQTarazonaEvans2global} read
as
\begin{align}
\int d1 V_\text{ext}(1) \dotsc \int dn V_\text{ext}(n) (\rv_\alpha
\times \nabla_\alpha + \ov_\alpha \times \nabla_\alpha^\omega )H_n =
0
\end{align} 
and 
\begin{align}\int d 1 \rho(1) \dotsc \int d n \rho(n) (\rv_\alpha \times
\nabla_\alpha + \ov_\alpha \times \nabla_\alpha^\omega )c_n = 0.
\end{align}
Here we have used the shorthand notation $1\equiv \textbf{r}_1,
\boldsymbol{\omega}_1$ etc., and the derivation is analogous to
the above rotational case of spherical particles. 

The two-body sum rules \eqref{EQTarazonaEvans1} and
\eqref{EQTarazonaEvans2} are identical to those obtained by Tarazona
and Evans \cite{tarazona1983} 
using rotational
invariance arguments applied directly to correlation functions. Our
methodology not only allows to naturally re-derive their results, but
also to identify the full gamut of adiabatic rotatational sum rules,
from the global statements \eqref{EQTarazonaEvans1global} and
\eqref{EQTarazonaEvans2global} to the infinite hierarchies
\eqref{EQTarazonaEvans1hierarchy} and
\eqref{EQTarazonaEvans2hierarchy}.
(We use notational convention different
  from Tarazona and Evans~\cite{tarazona1983}: our $\nabla^\omega$ is (only)
  a partial derivative with respect to $\ov$,
  i.e.\ $\nabla^\omega\equiv\partial/\partial\ov$, whereas their
  $\nabla_\omega\equiv\ov\times\partial/\partial\ov$. The modulus is
  fixed, $|\ov|=1$, in both versions. Tarazona and Evans~\cite{tarazona1983} notate $H_2$ as $G$ in their\ (18) and (19).)

Again for each microstate $\sum_i (\rv_i\times\nabla_i u +
\ov_i\times\nabla_i^\omega u)=0$ and hence on average $\Torque_{\rm
  int}^{\rm tot}=0$. From the splitting $\Torque_{\rm int}^{\rm
  tot}=\Torque_{\rm ad}^{\rm tot}+\Torque_{\rm sup}^{\rm tot}$, we
conclude $\Torque_{\rm sup}^{\rm tot}=0$. From rotational invariance
of $P_t^{\rm exc}[\rho,\Jv,\Jv^\omega]$, where $\Jv^\omega(\rv,\ov,t)$
is the rotational current \cite{krinninger2016prl}, against an
instantaneous angular ``kick'' $\delta \Jv=(\dot\nv\times \rv) \rho$
and $\delta \Jv^\omega = (\dot\nv \times \ov) \rho$, we find 
\begin{align}
\Torque_{\rm sup}^{\rm tot}=-\int d\rv
d\ov\rho(\rv,\ov,t)(\rv\times\frac{\delta P_t^{\rm
    exc}}{\delta\Jv}+\ov\times\frac{\delta P_t^{\rm exc}}{\delta
  \Jv^\omega})=0.
\end{align}  
    Local sum rules can be obtained straightforwardly
by building derivatives with respect to $\Jv(\rv,\ov,t)$ and
$\Jv^\omega(\rv,\ov,t)$. The result is: 
\begin{align}
\int d\rv_1 d\ov_1\rho(1)
[\rv_1\times \delta/\delta
  \Jv(1)+\ov_1\times\delta/\delta\Jv^\omega(1)]{\sf M}_{n,m}=0. \label{EQMnm}
\end{align}  
 Here the tensorial equal-time direct correlation functions are defined as
${\sf M}_{n,m}=-\beta\delta^{n+m} P_t^{\rm
  exc}/\delta\Jv(1)\ldots\delta\Jv(n)\delta
\Jv^\omega(n+1)\ldots\Jv^\omega(n+m)$, where the roman numerals refer
to position, orientation and time $t$ (no index), e.g.\ $1\equiv
\rv_1,\ov_1, t$. We have assumed that all functional derivatives
commute.

So far we have restricted ourselves to uniaxial particles. 
The derived rotational sum rules can be analogously determined for general anisotropic particles with an additional orientation vector $\boldsymbol{\varpi}$.  
This can be done by simply replacing 
$\ov\times\nabla^\omega \to \ov\times\nabla^\omega + \boldsymbol{\varpi}\times\nabla^\varpi$, $\ov \times \frac{\partial}{\partial \textbf{J}^\omega} \to \ov \times \frac{\partial}{\partial \textbf{J}^\omega}+\boldsymbol{\varpi} \times \frac{\partial}{\partial \textbf{J}^\varpi}$ and $\int d \textbf{r} d\ov \to \int d \textbf{r} d\ov  d \boldsymbol{\varpi}$
in \eqref{EQTarazonaEvans1global}-\eqref{EQMnm}. 
This replacement also affects the adiabatic spin torque field $\tau_\text{ad}$ and all one-body field depend additionally to $\textbf{r}$ and $\ov$ on the orientation $\boldsymbol{\varpi}$.
For anisotropic particles there exists a more general version of \eqref{EQMnm} by exchange of ${\sf M}_{n,m} \to {\sf M}_{n,m,l}$, where $l$ denotes the number of functional derivations with respect to the rotational current $\textbf{J}^\varpi$. The indices $n$ and $m$ belong to the number of functional derivatives with respect to $\textbf{J}$ and $\textbf{J}^\omega$ as before.

\mysubsec{Memory invariance}{SECmemoryInvariance}In the above treatment
of nonequilibrium situtations we have exploited invariance against an
instantaneous transformation applied to the system. As we have shown, the corresponding
Noether identities carry imminent physical meaning. These sum rules
hold for the nonequilibrium effects that arise from the interparticle
interaction, i.e.\ they constrain superadiabatic forces and torques,
as obtained from translation and rotation, respectively. Here we
exploit that the corresponding nonequilibrium functional generator
$P_t^{\rm exc}$ carries further invariances, once one allows the
transformation to act also on the history of the system. As we
demonstrate in the following, the resulting identities constitute
exact constraints on the memory structure that are induced by the
coupled interparticle interactions. Recall that a reduced one-body
description of a many-body system is generically non-Markovian
(i.e.\ nonlocal in time) \cite{zwanzig2001}. The study of memory
kernels, often carried out in the framework of generalized Langevin
equations, is a topic of significant current research activity
\cite{lesnicki2016,lesnicki2017,jung2016,jung2017, jung2018,
  yeomansreyna2000, chavezrojo2006, lazarolazaro2017}.

Our approach differs from these efforts in that no a priori generic
form of a reduced equation of motion is assumed. Rather our
considerations are formally exact and interrelate (and hence
constrain) time correlation functions, which are generated from the
central nonequilibrium object $P_t^{\rm exc}$ via functional
differentiation. Very little is known about the memory structure of
superadiabatic forces, with exceptions being the NOZ framework
\cite{brader2013noz,brader2014noz} and the demonstration of the
relevance of memory for the observed viscoelasticity of hard sphere
liquids \cite{treffenstaedt2020shear}.
Both the Ornstein-Zernike (OZ) and NOZ relations are different from the Noether identities. The former relations are a direct consequence of the generality of the variational principle. Per se, neither the OZ nor the NOZ relations reflect the Noether symmetries.

Recalling the illustrated overview of the different types of shifting
in Fig.~\ref{fig1}, in the following we treat two further types of
invariance transformations: One is the static transformation.
This operation is formally analogous to the above equilibrium
treatment of the adiabatic state, but it is here carried out in the
same way at all times.  This static transformation contrasts
(and complements) the instantaneous transformation used above
for the time-dependent case. The corresponding changes to density and
current are graphically illustrated in Fig.~\ref{fig3}(b). The second
invariance operation is that of memory shifting, where the
transformation parameter is taken to be time-dependent,
cf.\ Fig.~\ref{fig3}(c) for a graphical representation. For simplicity
we restrict ourselves to cases where at both ends of the considered
time interval, no shifting occurs (i.e.\ such that the transformation
is the identity at the limiting times).

The static spatial shift consists of
$\rho(\rv,t')\to\rho(\rv+\eps,t')$ and
$\Jv(\rv,t')\to\Jv(\rv+\eps,t')$, where the time argument $t'$ is
arbitrary, and $\eps=\rm const$ characterizes magnitude and direction
of the translation, see the illustration in Fig.~\ref{fig3}(b). Hence
the time derivative $\dot\eps=0$ such that the current does not
acquire any displacement contribution, as also $-\dot\eps\rho=0$ at
all times. The spatial shift applies to all times $t'$ considered, and
we can restrict ourselves to $0\leq t'\leq t$.  Hence the changes in
kinematic fields are to first order given by
$\delta\rho(\rv,t')=\eps\cdot\nabla\rho(\rv,t')$ and
$\delta\Jv(\rv,t')=\eps\cdot\nabla\Jv(\rv,t')$.
As $P_t^{\rm exc}$ originates solely from the interparticle
interaction potential, the invariance of $u(\rv^N)$ against the global
displacement at all times induces invariance of $P_t^{\rm exc}$. Hence
$P_t^{\rm exc}[\rho,\Jv]=P_t^{\rm
  exc}[\rho+\delta\rho,\Jv+\delta\Jv]=P_t^{\rm exc}[\rho,\Jv]+ \delta
P_t^{\rm exc}$. Here $\delta P_t^{\rm exc}$ indicates the change in
superadiabatic free power and due to the invariance $\delta P_t^{\rm
  exc}=0$. On the other hand we can express $\delta P_t^{\rm exc}$ via
the functional Taylor expansion. To linear order the result consists
of two integrals.  One integral comes from the time-slice functional
derivative at fixed (end) time~$t$ and is given by $\int d\rv [
  (\delta P_t^{\rm exc}/\delta\rho)\delta\rho +(\delta P_t^{\rm
    exc}/\delta\Jv)\cdot \delta \Jv]$.  The second integral is from a
functional derivative at (variable) time $t'$, given by $ \int_0^t
dt'\int d\rv' [ (\delta P_t^{\rm exc}/\delta\rho')\delta\rho' +(\delta
  P_t^{\rm exc}/\delta\Jv')\cdot\delta\Jv' ]$, where the prime
indicates dependence on arguments $\rv'$ and $t'$. We then exploit
that the displacement $\eps$ of the static shift (which parametrizes
the changes in density and in current) is arbitrary. The result is a
global nonequilibrium Noether Theorem, given by
\begin{align}
  &  \int d\rv \Big(
  \frac{\delta P_t^{\rm exc}}{\delta\rho}\nabla\rho
  -\fv_{\rm sup}\cdot \nabla\Jv^{\sf T}\Big)
  \label{EQstaticShiftNoether}
  \\  &\quad 
  +\int_0^t dt'\int d\rv' \Big(
  \frac{\delta P_t^{\rm exc}}{\delta\rho'}\nabla'\rho'
  +\frac{\delta P_t^{\rm exc}}{\delta\Jv'}\cdot\nabla'\Jv'^{\sf T}
  \Big)
  =0, \notag
\end{align}
where we have used the relationship of the superadiabatic force field
to its generator, $\fv_{\rm sup}(\rv,t)=-\delta P_t^{\rm
  exc}[\rho,\Jv]/\delta \Jv(\rv,t)$, the primed symbol $\nabla'$
indicates the derivative with respect to $\rv'$, and the superscript
$\sf T$ denotes the matrix transpose. (In index notation the
$k$-component of the vector ${\bf a}\cdot\nabla {\bf b}^{\sf T}$ is
$\sum_{k'} a_{k'} \nabla_k b_{k'}$.)

Equation \eqref{EQstaticShiftNoether} constitutes a global identity
that links density, current, and superadiabatic force field in a
nontrivial spatial and temporal form. As the central variation
principle \cite{schmidt2013pft} allows to vary $\Jv(\rv,t)$ freely,
\eqref{EQstaticShiftNoether} remains true upon building the functional
derivative with respect to $\Jv(\rv,t)$. The result is a local
identity
\begin{align}
&  \beta\nabla \fv_{\rm sup}=\notag \\&
  \int d\rv' \big( \mv_2(\rv,\rv',t)\nabla'\rho(\rv',t)
  + {\sf M}_2(\rv,\rv',t)\cdot\nabla'\Jv^{\sf T}(\rv',t)
  \big)
  \notag\\ & \quad
  +\int_0^t dt' \int d\rv'\big(
  \mv_2(\rv,t,\rv',t') \nabla'\rho'
  \label{EQstaticShiftingGradfsup}
  \\ & \qquad\qquad\qquad\qquad
  + {\sf M}_2(\rv,t,\rv',t')\cdot\nabla'\Jv'^{\sf T}
  \big),\notag
\end{align}
where two-body time direct correlation functions occur in vectorial
form: $\mv_2(\rv,\rv',t)=-\beta\delta^2 P_t^{\rm
  exc}/\delta\Jv(\rv,t)\delta\rho(\rv',t)$,
$\mv_2(\rv,t,\rv',t')=-\beta\delta^2 P_t^{\rm exc}/
\delta\Jv(\rv,t)\delta\rho(\rv',t')$, as well as in tensorial form:
${\sf M}_2(\rv,\rv',t)=-\beta\delta^2P_t^{\rm
  exc}/\delta\Jv(\rv,t)\delta\Jv(\rv',t)$,
${\sf M}_2(\rv,t,\rv',t')=-\beta\delta^2 P_t^{\rm
  exc}/\delta\Jv(\rv,t)\delta\Jv(\rv',t')$.
Here we have made the (common) assumption that the second derivatives
can be interchanged.

Repeated differentiation of \eqref{EQstaticShiftingGradfsup} with
respect to $\Jv(\rv,t)$ generates a hierarchy,
\begin{align}
  &  \sum_{\alpha=1}^{n-1} 
  \nabla_\alpha {\sf M}_{n-1}(\rv^{n-1},t)=
  \\
  &\quad  \int d\rv_n \Big( \mv_n(\rv^n,t) \nabla_n \rho(\rv_n,t)
  +{\sf M}_n(\rv^n,t)\cdot\nabla_n\Jv(\rv_n,t)^{\sf T}
  \Big)  \notag\\
  &\quad + \int_0^t dt'\int d\rv_n\Big(
  \mv_n(\rv^{n-1},t,\rv_n,t')
  \nabla_n\rho(\rv_n,t')\notag\\
  & \qquad\qquad\qquad\qquad
  +{\sf M}_n(\rv^{n-1},t,\rv_n,t')\cdot\nabla_n\Jv(\rv_n,t')^{\sf T}
  \Big)\notag,
\end{align}
where the $n$-body equal-time direct correlation functions of rank $n$
are ${\sf M}_{n}(\rv^n,t)=-\beta\delta^n P_t^{\rm exc}/\delta
\Jv(\rv_1,t)\ldots\delta\Jv(\rv_n,t)$ and
  $\mv_n(\rv^n,t)=-\beta\delta^n P_t^{\rm
    exc}/\delta\Jv(\rv_1,t)\ldots\delta\Jv(\rv_{n-1},t)\delta\rho(\rv_n,t)$,
where we have used the shorthand $\rv^n=\rv_1\ldots\rv_n$. Furthermore
at unequal times we have:
${\sf M}_{n}(\rv^{n-1},t,\rv_n,t')=-\beta\delta^n P_t^{\rm exc}/\delta
\Jv(\rv_1,t)\ldots\delta\Jv(\rv_{n-1},t) \delta\Jv(\rv_n,t')$ as a
rank $n$ tensor, and also a rank $n-1$ tensor
$\mv_n(\rv^{n-1},t,\rv_n,t')=-\beta \delta^n P_t^{\rm exc}/\delta
\Jv(\rv_1,t)\ldots\delta \Jv(\rv_{n-1},t)\delta\rho(\rv_n,t')$.

In the second case, we consider a more general invariance
transformation that is obtained by letting the transformation
parameter be time-dependent. In this case of time-dependent shifting,
we prescribe a displacement vector $\eps(t')$ for times $0\leq t'\leq
t$, i.e.\ between the intial time, throughout the past and up to the
``current'' time $t$. We restrict ourselves to vanishing shift at the
boundaries of the considered time interval,
i.e.\ $\eps(0)=\eps(t)=0$. Due to the overdamped character of the
dynamics, its interparticle contributions are unaffected by this
transformation, and hence $P_t^{\rm exc}$ is invariant. The induced
changes that the density and the current acquire arise from shifting
their position argument, but the current also acquires an additive
shifting current contribution. The latter contribution is
analogous to the (sole) effect that is present in the instantaneous
shifting, but here applicable at all times (in
the considered time interval).

Hence the time-dependent shifting, as illustrated in
Fig.~\ref{fig3}(c), induces the following changes to the density and
the current: $\delta\rho(\rv',t')=\eps(t')\cdot\nabla'\rho(\rv',t')$
and
$\delta\Jv(\rv',t')=\eps(t')\cdot\nabla\Jv(\rv',t')-\dot\eps(t')\rho(\rv',t')$,
where $\dot\eps(t')=d\eps(t')/dt'$. Next we can regard $P_t^{\rm
  exc}[\rho+\delta\rho,\Jv+\delta\Jv]$ as a functional of $\eps(t')$
and $\dot\eps(t')$. Its invariance amounts to stationarity,
i.e.\ vanishing first functional derivative, with respect to the
displacement. This problem, in particular for the present case of
fixed boundary values, amounts to one of the most basic problems in
the calculus of variations. It is realized, e.g., in the determination
of catenary curves and indeed, in Hamilton's principle of classical
mechanics. Exploiting the corresponding Euler-Lagrange equation leads
to
\begin{align}
  \frac{d}{dt'} \int d\rv' \mv'_1 \rho' 
  + \int d\rv' \mv_1'\cdot\nabla'\Jv'^{\sf T}
  +\int d\rv' m_1'\nabla'\rho' &= 0,
  \label{EQfinal}
\end{align}
where the one-body time direct correlation functions are
$\mv'_1(\rv',t',t)=-\beta\delta P_t^{\rm exc}/\delta\Jv(\rv',t')$ and
$m'_1(\rv',t',t)=-\beta\delta P_t^{\rm exc}/\delta\rho(\rv',t')$.
Differentiation with respect to $\Jv(\rv,t)$ yields again a local
memory identity.

\vspace{5mm}
\mysubsec{Conclusions}{SECconclusions}We have demonstrated that Noether's
Theorem for exploiting symmetry in a variational context has profound
implications for Statistical Physics. Known sum rules can be derived
with ease and powerfully generalized to full infinite hierarchies, to
the rotational case, and to time-dependence in nonequilibrium. 
Recall the selected applications \cite{xu2008, brader2008precursor, walz2010, haering2015, haering2020phd, bryk1997, tejero1985, iatsevitch1997, kasch1993, henderson1984, mandal2017} of the equilibrium sum 
rules, as we have laid out in the introduction. For the time-dependent
case, we envisage similar insights from using the newly formulated 
nonequilibrium sum rules in investigations of e.g\ the dynamics of 
freezing, of liquid crystal flow, and of driven fluid interfaces.
On the conceptual level, Noether's Theorem assigns a clear meaning 
and physical interpretation to all resulting identities, as being
generated from an invariance property of an underlying functional
generator. Although the symmetry operation that we considered are
simplistic, and only their lowest order in a power series expansion
needs to be taken into account, a significantly complex body of sum
rules naturally emerges. Hence the application of Noether's Theorem is
significantly deeper than mere exploiting of symmetries of arguments
of correlation functions, i.e.\ that the direct correlation function
$c_2(\rv,\rv')$ in bulk fluids depends solely on $|\rv-\rv'|$. Rather,
as governed by functional calculus, coupling of different levels of
correlation functions occurs.

Crucially, the Noether sum rules are different from the variational principle, as embodied in the Euler-Lagrange equation. On the formal level, the difference is that the Euler-Lagrange equation (both of DFT and of PFT) is a formally closed equation on the one-body level. In contrast, the Noether rules couple $n$- and $(n+1)$-body correlation functions, hence they are of genuine hierarchical nature. They also describe different physics, as the Euler-Lagrange equation expresses a chemical potential equilibrium in DFT and the local force balance relationship in PFT. In contrast, the Noether identities stem from the symmetry properties of the respective underlying physical system.

The standard DFT approximations, ranging from simple local, square-gradient, and mean-field functional to more sophisticated weighted-density-schemes including fundamental measure theory satisfy the internal force relationships. This can be seen straightforwardly by observing that these functionals do satisfy global translation invariance. 
(The value of the free energy is independent of the choice of coordinate origin.) All higher-order Noether identities are then automatically satisfied, as these inherit the correct symmetry properties from the generating (excess free energy) functional. Our formalism hence provides a concrete reason, over mere empirical experience, why the practitioners' choices for approximate functionals are sound. The situation for more complex DFT schemes could potentially be different though. As soon as, say, self-consistency of some form is imposed, or coupling to auxiliary field comes into play, it is easy to imagine that the Noether identities help in restricting choices in the construction of such approximation schemes.

The sum rules imposed by the three types of dynamical displacements are satisfied within the velocity gradient form of the power functional \cite{delasheras2018velocityGradient,delasheras2020prl}. It is straightforward to see that the functional is independent of the coordinate origin (static shifting). For the cases of dynamical shifting, the invariance of the functional stems from invariance of the velocity field against shifting. For both instantaneous and memory shifting, the velocity gradient remains invariant under the displacement.

We envisage that the higher than two-body Noether identities can facilitate the construction of advanced liquid state/density functional approximations. Such work should surely be highly challenging. In the context of fundamental measure theory (see e.g.\ the work by Roth \cite{roth2010review} for an enlightening review) it is worth recalling that in Rosenfeld's original 1989 paper \cite{rosenfeld1989}, he calculated the three-body direct correlation function from his then newly proposed functional. The result for the corresponding three-body pair correlations compared favourably against simulation data. Furthermore, the recent insights into two-body correlations in inhomogeneous liquids \cite{tschopp2021} and crystals \cite{oettel2021} demonstrates that working with higher-body correlation functions is feasible.

In future work it would be very beneficial to bring together the
Noether identities with the nonequilibrium Ornstein-Zernike relations
\cite{brader2013noz,brader2014noz}, in order to aid construction of
new dynamical approximations. One could exploit the rotational
invariance of the superadiabatic excess power functional $P_t^{\rm
  exc}$ to gain deeper insights in its memory structure and also
generalize the translational memory relations 
 \eqref{EQstaticShiftNoether}-\eqref{EQfinal} to anisotropic particles.
It would be highly interesting to apply \eqref{EQTZ} to the recently obtained direct correlation function of the hard sphere crystal. This would allow to investigate whether Triezenberg and Zwanzig's concept that they originally developed for the free gas-liquid interface applies to the also self-sustained density inhomogeneity in a solid.
  Furthermore,
addressing further cases of self motility
\cite{farage2015,paliwal2018,paliwal2017activeLJ},  including active freezing \cite{turci2020,geissler2021}, as well as further
types of time evolution,  
such as molecular dynamics or quantum
mechanics should be interesting.  This is feasible, as the Noether
considerations are not restricted to overdamped classical systems, as
(formal) power functional generators exist for quantum
\cite{schmidt2015qm} and classical Hamiltonian \cite{schmidt2018md}
many-body systems.
On the methodological side, besides power functional theory, our 
framework could be complemented by e.g.\ mode-coupling theory and  
Mori-Zwanzig techniques \cite{janssen2018}, as well as approaches 
beyond that \cite{fuchs2020}.
Given that the equilibrium force sum rules are crucial in the
description of crystal \cite{walz2010,haering2015} and liquid crystal
\cite{haering2020phd} excitations, the study of such systems under
drive is a further exciting prospect.

\mysec{Methods}{SECMethods}
\mysubsec{Relationship to classical results}{SECOverview}
We give an overview of how the Noether sum rules relate to previously known results. 
The famous LMBW-equation was derived independently by Lovett, Mou, and  Buff \cite{lovett1976} and by Wertheim \cite{wertheim1976} and reads 
\begin{align}
\nabla \ln \rho(\textbf{r}) + \beta \nabla V_\text{ext}(\textbf{r}) = \int d \textbf{r}' c_2(\textbf{r},\textbf{r}') \nabla' \rho(\textbf{r}'). \label{EQLMBW}
\end{align}
We can conclude that \eqref{EQLMBW} is a combination of the local internal Noether sum rule  \eqref{EQBobDirectTwoBody} for translational symmetry and the equilibrium Euler-Lagrange equation $c_1(\textbf{r}) = \ln \rho(\textbf{r}) + \beta \nabla V_\text{ext}(\textbf{r}) - \beta \mu$, where $\mu$ indicates the chemical potential.
LMBW also derived a lesser known external relation, which is equivalent to \eqref{EQLMBW} and reads 
\begin{align}
&\nabla \rho(\textbf{r}) + \beta \rho(\textbf{r}) \nabla V_\text{ext}(\textbf{r}) = \nonumber\\
&\qquad\int d \textbf{r}' (g(\textbf{r},\textbf{r}') -1) \rho(\textbf{r}) \rho(\textbf{r}') \nabla' V_\text{ext}(\textbf{r}'). \label{EQLMBW2}
\end{align}
We find that \eqref{EQLMBW2} contains the local external Noether sum rule \eqref{EQBobH2} along with the relation $H_2(\textbf{r},\textbf{r}') = (g(\textbf{r},\textbf{r}') -1) \rho(\textbf{r}) \rho'(\textbf{r}') + \rho(\textbf{r}) \delta(\textbf{r}-\textbf{r}')$ \cite{hansen2013}.

The Triezenberg-Zwanzig equation \cite{TZ1972} holds for vanishing external potential $V_\text{ext}(\textbf{r})=0$ and is given by
\begin{align}
\nabla \ln \rho(\textbf{r}) = \int d \textbf{r}' c_2(\textbf{r},\textbf{r}') \nabla' \rho(\textbf{r}'). \label{EQTZ}
\end{align}
Originally \eqref{EQTZ} was derived for the free liquid-vapour interface in a parallel geometry. We find that the relation consists of the local internal Noether sum rule  \eqref{EQBobDirectTwoBody} and the equilibrium Euler-Lagrange equation. The LMBW equation \eqref{EQLMBW2} reduces to the Triezenberg-Zwanzig equation \eqref{EQTZ} for cases of $V_\text{ext}(\textbf{r})=0$.

Another related equation is the first member of  the Yvon-Born-Green (YBG) hierarchy \cite{yvon,borngreen},
\begin{align}
&\rho( \nabla \ln \rho(\textbf{r}) + \beta \nabla V_\text{ext}(\textbf{r})) = \nonumber \\
&\qquad - \beta \int d \textbf{r}' g(\textbf{r},\textbf{r}') \rho(\textbf{r}) \rho(\textbf{r}') \nabla \phi(|\textbf{r}-\textbf{r}'|), \label{EQYBG}
\end{align}
where $\phi(r)$ denotes the interparticle pair potential as before.
Although \eqref{EQYBG} has a similar structure as the LMBW equation, it is not based on symmetry or Noether arguments but arises from integration out of degrees of freedom. If one would like to include the translational symmetry one can simply replace the left hand side of \eqref{EQYBG} with the right hand side of the LMBW equation \eqref{EQLMBW}, which leads to
\begin{align}
\int d \textbf{r}' c_2(\textbf{r},\textbf{r}') \nabla' \rho(\textbf{r}')= - \beta \int d \textbf{r}' g(\textbf{r},\textbf{r}') \rho(\textbf{r}') \nabla \phi(|\textbf{r}-\textbf{r}'|).
\end{align}

Some of the here derived sum rules are rederivations of known relations. We reiterate the relationships. In his overview Baus \cite{baus1984} showed that in equilibrium the total external \eqref{EQforceExternalTotal} and internal \eqref{EQforceInternalTotal} force vanish and derived the corresponding local hierarchies \eqref{EQBobH2}, \eqref{EQBobHnHierarchy} and \eqref{EQBobDirectTwoBody}, \eqref{EQBobDirectHierarchy}. Similar sum rules  \cite{baus1984} hold for the external 
\eqref{EQTorqueExtTotVanishes} and internal \eqref{EQTorqueAdiabaticTotVanishes} total torques and their corresponding hierarchies \eqref{EQtorqueH2}, \eqref{EQtorqueHn} and \eqref{EQtorquec2}, \eqref{EQtorquecn}.
Tarazona and Evans \cite{tarazona1983} generalized these equations for uniaxial particles and derived the first order of the external \eqref{EQTarazonaEvans1} and internal \eqref{EQTarazonaEvans2} hierarchies due to rotations.

To the best of our knowledge the hierarchies of global Noether sum rules, such as \eqref{EQHn}, \eqref{EQcn}, \eqref{EQHnrot}, and  \eqref{EQcnrot}, have not been determined previously. 
Furthermore the global external \eqref{EQTarazonaEvans1global} and internal \eqref{EQTarazonaEvans2global} Noether sum rules for uniaxial colloids and their corresponding hierarchies \eqref{EQTarazonaEvans1hierarchy} and \eqref{EQTarazonaEvans2hierarchy} (with exception of the first order) are reported here for the first time.
As the considerations in the literature focused on equilibrium, all our nonequilibrium relations, as e.g.\ \eqref{EQforceSuperadiabaticTotal}-\eqref{EQMnidentity} and especially the ones including memory \eqref{EQstaticShiftNoether}-\eqref{EQfinal} have not been found before.

\mysec{Acknowledgements}{SECAcknowledgements} 
This work is supported by the German Research Foundation (DFG) via project number 436306241. 

We would like to thank Matthias Fuchs, Johannes H{\"a}ring, Bob Evans, Pedro Tarazona, Daniel de las Heras, Matthias Kr{\"u}ger, Jim Lutsko, Hartmut L{\"o}wen, and Thomas Fischer for useful discussions and comments.

\end{document}